\documentclass{aastex631}

\usepackage{xcolor}

\shorttitle{X-ray performance of critical-angle transmission grating prototypes for the Arcus mission}
\shortauthors{Heilmann et al.}
\graphicspath{{./}{figures/}}

\begin{document}

\title{X-ray performance of critical-angle transmission grating prototypes for the Arcus mission}

\author[0000-0001-9980-5295]{Ralf K. Heilmann}
\affiliation{Space Nanotechnology Laboratory \\
MIT Kavli Institute for Astrophysics and Space Research \\
Cambridge, MA 02139, USA}
\author{Alexander R. Bruccoleri}
\affiliation{Izentis LLC, Cambridge, MA 02139, USA}
\author{Vadim Burwitz}
\affiliation{Max-Planck-Institut f\"ur Extraterrestrische Physik, 85748 Garching, Germany}
\author[0000-0002-9184-4561]{Casey deRoo}
\affiliation{University of Iowa, Iowa City, IA 52242, USA}
\author{Alan Garner}
\affiliation{MIT Kavli Institute for Astrophysics and Space Research \\
Cambridge, MA 02139, USA}
\author[0000-0003-4243-2840]{Hans Moritz G\"unther}
\affiliation{MIT Kavli Institute for Astrophysics and Space Research \\
Cambridge, MA 02139, USA}
\author{Eric M. Gullikson}
\affiliation{Lawrence Berkeley National Laboratory, Berkeley, CA 94720, USA}
\author{Gisela Hartner}
\affiliation{Max-Planck-Institut f\"ur Extraterrestrische Physik, 85748 Garching, Germany}
\author{Ed Hertz}
\affiliation{Center for Astrophysics, Harvard-Smithsonian Astrophysical Observatory, Cambridge, MA 02138, USA}
\author{Andreas Langmeier}
\affiliation{Max-Planck-Institut f\"ur Extraterrestrische Physik, 85748 Garching, Germany}
\author[0000-0002-0717-0462]{Thomas M\"uller}
\affiliation{Max-Planck-Institut f\"ur Extraterrestrische Physik, 85748 Garching, Germany}
\author{Surangkhana Rukdee}
\affiliation{Max-Planck-Institut f\"ur Extraterrestrische Physik, 85748 Garching, Germany}
\author{Thomas Schmidt}
\affiliation{Max-Planck-Institut f\"ur Extraterrestrische Physik, 85748 Garching, Germany}
\author[0000-0003-4284-4167]{Randall K. Smith}
\affiliation{Center for Astrophysics, Harvard-Smithsonian Astrophysical Observatory, Cambridge, MA 02138, USA}
\author[0000-0001-6932-2612]{Mark L. Schattenburg}
\affiliation{Space Nanotechnology Laboratory \\
MIT Kavli Institute for Astrophysics and Space Research \\
Cambridge, MA 02139, USA}



\begin{abstract}

Arcus is a proposed soft x-ray grating spectrometer Explorer.  It aims to explore cosmic feedback by mapping hot gases within and between galaxies and galaxy clusters and characterizing jets and winds from supermassive black holes, and to investigate the dynamics of protoplanetary discs and stellar accretion.  Arcus features 12 m-focal-length grazing-incidence silicon pore optics (SPO) developed for the Athena mission.  Critical-angle transmission (CAT) gratings efficiently disperse high diffraction orders onto CCDs.  
We report new and improved x-ray performance results for Arcus-like CAT gratings, including record resolving power for two co-aligned CAT gratings. 
Multiple Arcus prototype grating facets were illuminated by an SPO at the PANTER facility.  The facets consist of $32\times32.5$ mm$^2$ patterned silicon membranes, bonded to metal frames. The bonding angle is adjusted according to the measured average tilt angle of the grating bars in the membrane. Two simultaneously illuminated facets show minor broadening of the Al-K$_{\alpha}$ doublet in 18$^{\rm th}$ and 21$^{\rm st}$ orders with a best fit record effective resolving power of $R_G \approx 1.3^{+\infty}_{-0.5}\times10^4$ ($3\sigma$), about 3-4 times the Arcus requirement.  
We measured the diffraction efficiency of quasi-fully illuminated gratings at O-K wavelengths in orders 4-7 in an Arcus-like configuration and compare results with synchrotron spot measurements. After corrections for geometrical effects and bremsstrahlung continuum we find agreement between full and spot illumination at the two different facilities, as well as with the models used for Arcus effective area predictions.  We find that these flight-like gratings meet diffraction efficiency and greatly exceed resolving power Arcus requirements.

\end{abstract}



\section{Introduction} \label{sec:intro}

Spectroscopy lies at the roots of astrophysics.  High-resolution absorption and emission line spectroscopy in the soft x-ray band informs us about the physical conditions and chemical composition of warm and hot plasmas in the interstellar, circumgalactic and intergalactic media, and their roles in star and galaxy formation and cosmic feedback that controls the in- and outflows of galaxies and galaxy clusters. Magnetically active young stars, evolved coronal stars, and stellar accretion processes also emit soft x rays, giving us an opportunity to better understand the formation and evolution of stellar systems.

Resolving weak lines requires high spectrometer effective area $A_{\rm eff}$ to reduce statistical uncertainties, but also high spectral resolving power $R = \lambda/\Delta \lambda$ ($\lambda$ being the x-ray wavelength and $\Delta \lambda$ the smallest resolvable difference in $\lambda$) to distinguish a weak line from a surrounding continuum.
Most of our current astrophysical soft x-ray spectroscopy data stems from two instruments: The High Energy Transmission Grating Spectrometer (HETGS) on Chandra (\citep{cxc}), and the Reflection Grating Spectrometer (RGS) on XMM-Newton (\citep{RGS}).  Both missions were launched in 1999 with technology developed a generation ago.  Resolving power  for the HETGS can approach $1000$, but with rapidly diminishing effective area ($< 10$ cm$^2$ below 1 keV) toward low energies.  The RGS provides effective area up to about 100 cm$^2$, but with typical R only around 200.  Both aging instruments have provided tantalizing hints for the presence of hot ($10^7$ K) baryons in the extended halos of galaxies and clusters, for example, with a handful of reported absorption line detections at borderline statistical significance (\citep{Nicastro2018,Kov_cs_2019}). An x-ray spectrometer with improved resolving power and effective area similar to Arcus would facilitate a survey of these absorption systems (\citep{Bregman2015}).

The X-Ray Imaging and Spectroscopy Mission (XRISM) will fly an x-ray calorimeter spectrometer with $\sim 5$ eV energy resolution (\citep{XRISM_SPIE2021}), but below $\sim 1$ keV it will perform worse than the RGS.  Currently the only proven way to increase soft x-ray spectroscopy performance is via improved grating spectroscopy.

\begin{figure}[ht!]
   \begin{center}
   \begin{tabular}{ c c } 
   \includegraphics[height=4.5cm]{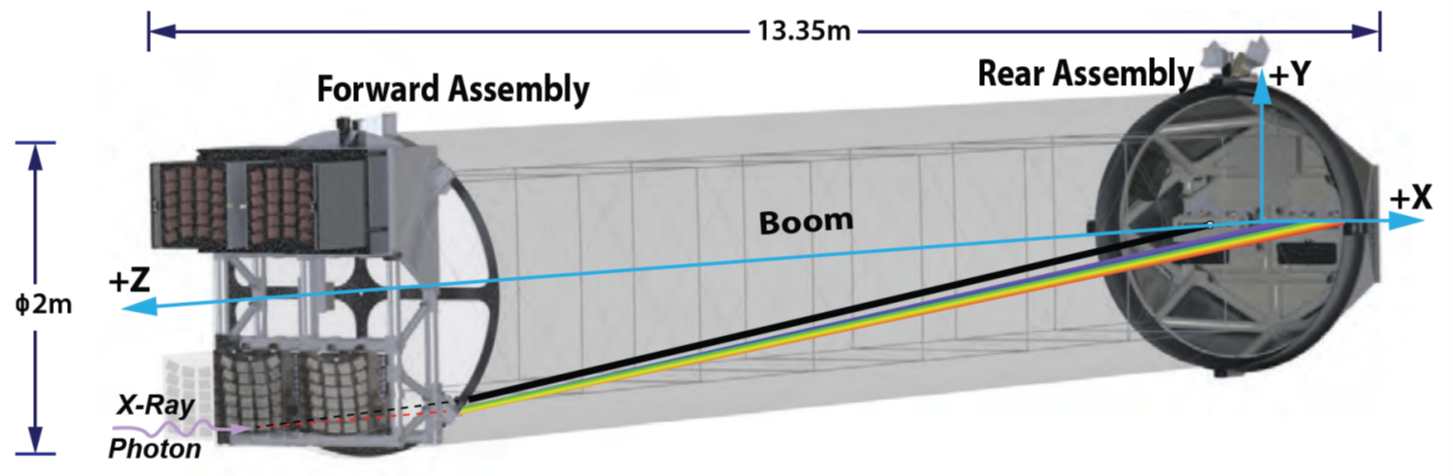}
   \includegraphics[height=4.5cm]{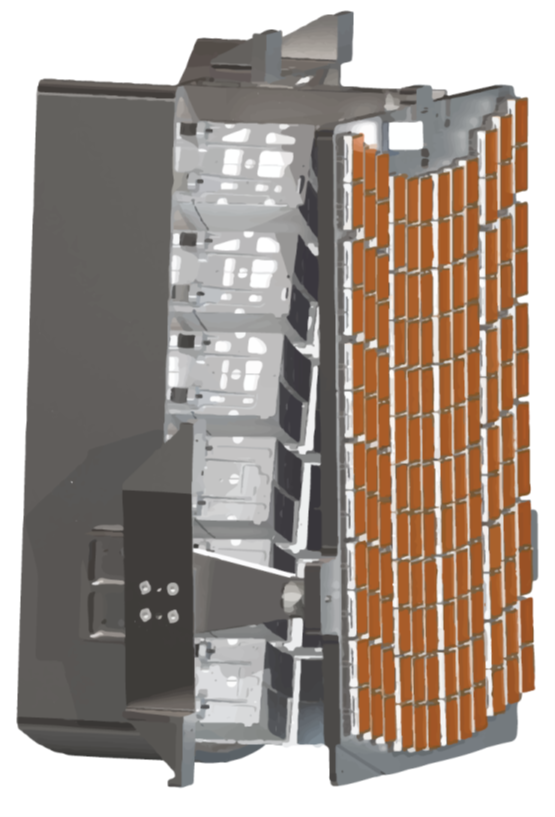}   
   \end{tabular}
   \end{center}
\caption{Design of the Arcus spectrometer instrument. Left: Overview with all four optical channels (OCs), showing from left to right: Precollimators (only shown for top two OCs), mirror/grating petals (only visible in bottom two OCs), the sock-encased deployable boom, 
and two linear CCD readout arrays at the rear.  Symbolically, photons are shown entering the lower left OC, with the black line showing the 0$^{\rm th}$ order (image) path leading to the left readout, and the rainbow-colored path representing the blazed dispersed photons leading to the right readout.  Right: Enlarged side view of one of four optics petals, comprised of an SPO mirror petal (left) and a CAT grating petal (right) (\citep{Arcus2020}). \label{fig:layout}}
\end{figure}

The Arcus grating spectrometer Explorer uses state-of-the-art technology and is designed to improve the spectroscopic figure-of-merit in the 12-50 \AA \ bandpass by a factor of 5-10 compared to HETGS and RGS, with a minimum R of 2500, and a goal of 3500, within the cost envelope of a NASA mid-size Explorer.  It uses four parallel optical channels (OC).  Each OC consists of a 12 m-focal-length x-ray telescope with an objective grating spectrometer.  Two CCD readout strips are shared between the four channels (\citep{Arcus2020}).  See Fig.~\ref{fig:layout} for details.

The keys to high spectral resolving power for a grating spectrometer are a telescope line-spread function (LSF) which is narrow in the dispersion direction and which is not degraded appreciably by grating diffraction, and diffracting as many incident photons as possible to the largest feasible diffraction angles, i.e., into the highest possible diffraction orders.  The key to large effective area is a large telescope effective area in combination with high grating diffraction efficiency.

Arcus uses grazing-incidence silicon pore x-ray optics (SPO) developed for ESA's Athena mission 
as focusing elements (\citep{Collon_SPIE2021}).  The telescope image and the dominant diffraction orders are collected by two CCID-94 back-illuminated CCD arrays (\citep{Arcus2020}). To maximize the FOM, Arcus employs arrays of 200 nm-period critical-angle transmission (CAT) gratings arranged along the surface of the tilted Rowland torus for each channel (\citep{moritz2017}).  CAT gratings are lightweight, alignment-insensitive, blazed transmission gratings with high diffraction efficiency in the soft x-ray band and high transparency for harder x rays. The latter property will provide Arcus with significant effective area between 150 and 800 cm$^2$ in 0$^{\rm th}$ order with CCD energy resolution ($\sim 70$ eV) in the  $\sim 1.5$-6 keV range.

CAT gratings have been under development for over a decade (\citep{OE2008}).  We have gradually increased absolute grating diffraction efficiency (from $\sim 15$\% to well over 30\%) (\citep{SPIE2021}) and  size (from a few to over 1000 mm$^2$) and improved grating uniformity (\citep{SPIE2017}).  Effective resolving power of individual gratings has been demonstrated at the $R_G \sim 10^4$ level with illuminated areas of $\sim 30-300$ mm$^2$ (\citep{SPIE2016,AO2019,SPIE2020}).  A linear array of four aligned gratings illuminated by a pair of SPOs previously showed $R \sim 3500$ (\citep{SPIE2018}).  Environmental testing (thermal cycling, vibration testing) did not degrade grating performance (\citep{SPIE2017}).  However, most of these previously tested gratings have been fabricated one by one in a hands-on approach not suitable for volume production (\citep{EIPBN2016}).

The focus of the present work is the testing of the x-ray performance of a recent set of flight-like prototype CAT gratings with Arcus dimensions in terms of resolving power and diffraction efficiency in Arcus-like configurations.  Grating fabrication was done on a set of tools compatible with the volume manufacturing of over 500 gratings that will be required for Arcus (\citep{SPIE2020}). 

In the following we first give a brief description of the CAT grating principle and its translation into a manufacturable device.  We then discuss grating-to-grating alignment using optical techniques in air and our experimental setup for x-ray measurements. Results for resolving power and diffraction efficiency are presented, and the latter are compared to synchrotron spot measurements and model predictions.  Finally, we discuss our results and summarize the work presented here.

\section{CAT grating principle, design and fabrication}

\begin{figure}[ht!]
    \gridline{
    \fig{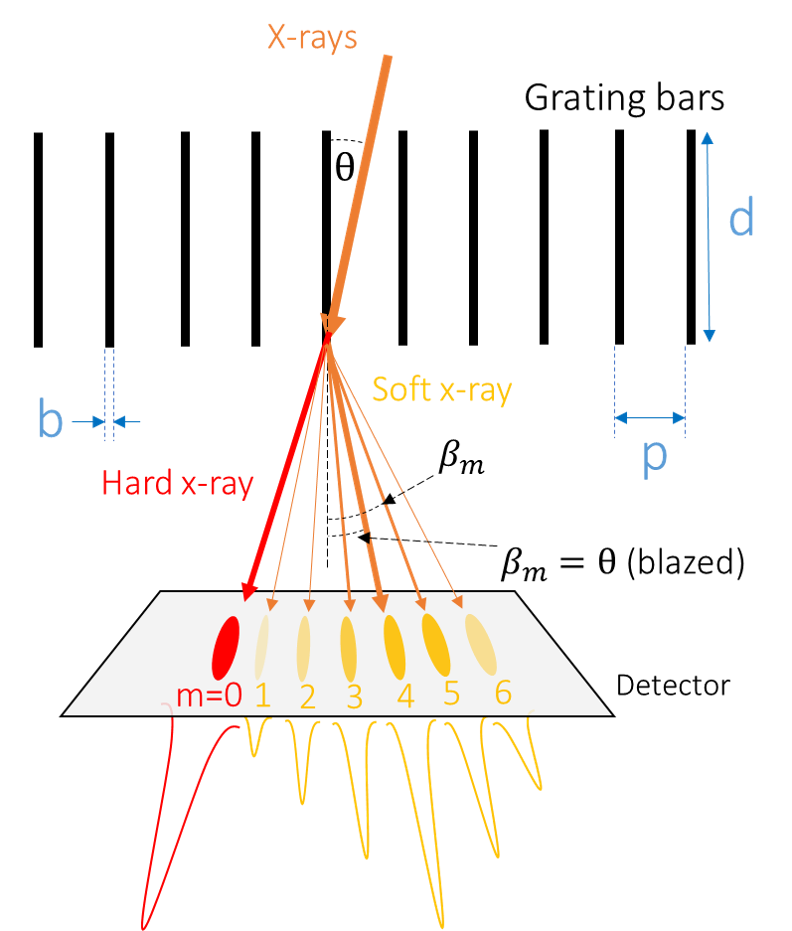}{0.4\textwidth}{(a)}
    \fig{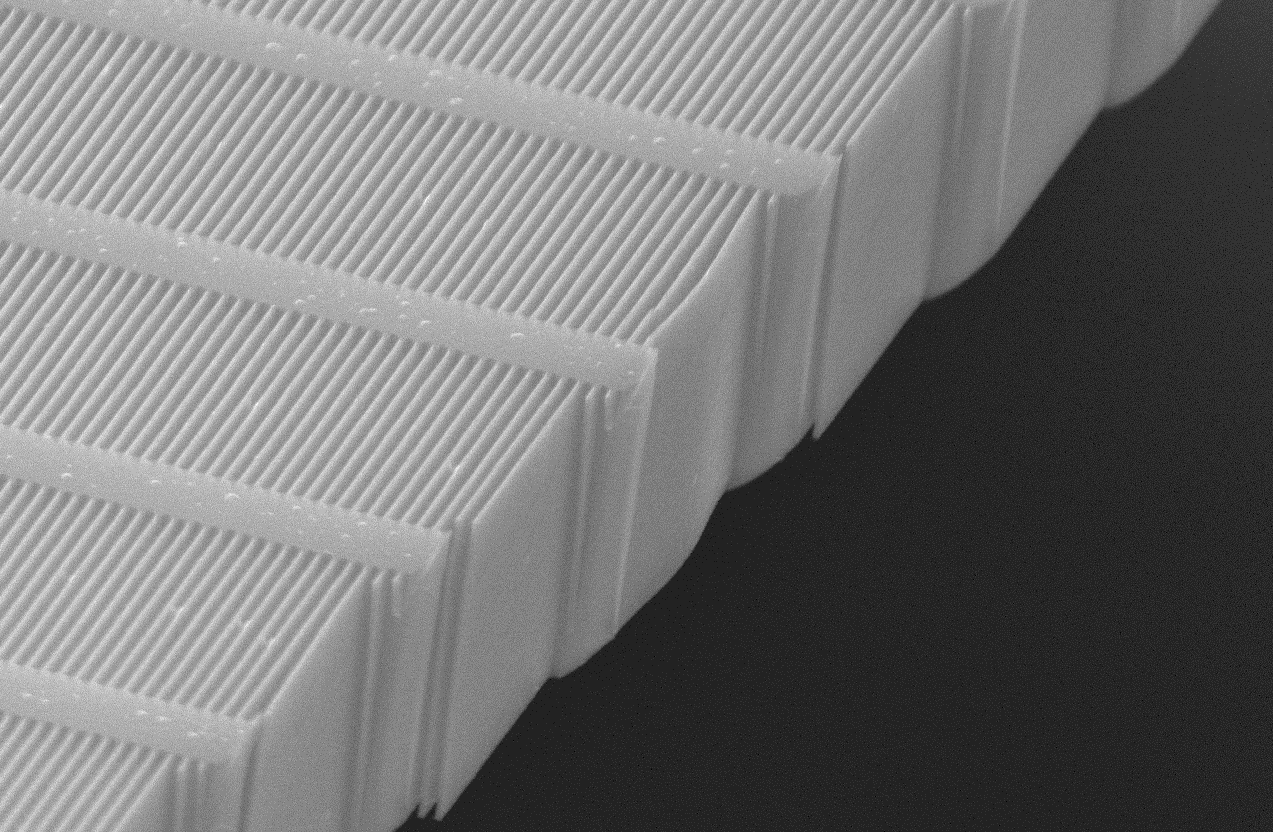}{0.4\textwidth}{(b)}
    }

\caption{(a) Schematic of CAT grating diffraction. X rays are incident at grazing angle $\theta$ onto the grating bar sidewalls. If $\theta < \theta_c(\lambda, n_r)$, efficient blazing enhances diffraction orders near the angle of specular reflection off the sidewalls ($m \sim 3-5$ in this schematic).  For harder x rays $\theta$ is above the critical angle, and most of the x rays get transmitted straight through the grating ($m = 0$). (b) Scanning electron micrograph (SEM) of a cleaved freestanding CAT grating.  CAT grating bars (200 nm period) and L1 cross supports (5 $\mu$m period) are seen from the top.  Smooth CAT grating bar sidewalls act as ``nanomirrors" at small angles of grazing incidence. \label{fig:CAT_schematic}}
\end{figure}

CAT gratings consist of freestanding grating bars with ultra-high aspect ratio $d/b$ (see Fig.~\ref{fig:CAT_schematic}).  Blazing is achieved by tilting the grating such that x rays impinge on the grating bar sidewalls at grazing angles of incidence $\theta$ below the critical angle for total external reflection, $\theta_c(\lambda, n_r)$, with $n_r$ being the index of refraction of the grating bar material.  If the sidewalls are smooth enough they effectively act as ``nanomirrors". This leads to enhancement (blazing) of the diffraction orders near the direction of specular reflection from the sidewalls.  
The grating equation provides the $m^{\mathrm {th}}$ order diffraction angle $\beta_m$ via

\begin{equation}
{m \lambda \over p} = \sin \theta - \sin \beta_m ,
\label{ge}
\end{equation}

\noindent
with $p$ being the grating period.  For soft x rays with energy $E < 1$ keV, a grating period of 200 nm, gratings made from silicon, and $\theta = 1.8^{\circ}$, one expects blazing to peak for orders around $m\lambda \approx 12-13$ nm, or $m \approx 3-10$, depending on wavelength.  Intuitively, efficiency in blazed orders can be maximized by making the grating bars as thin as possible and deep enough such that every ray that enters the gap between grating bars can be reflected in the specular direction and exit the grating without hitting another grating bar ($a/d = \tan \theta$).  Full electromagnetic modeling of CAT grating diffraction using rigorous coupled-wave analysis (RCWA, \citep{RCWA}) by-and-large supports this basic, ray-trace based design approach. While the CAT grating design promises high diffraction efficiency on the order of 50-60\%, it is challenging to fabricate freestanding grating bars with the required high aspect ratios and nm-smooth sidewalls.  Assuming a grating bar width of $b = 70$ nm and $\theta = 1.8^{\circ}$, for example, this requires $d \approx 4.1$ $\mu$m, and $d/b = 59$.

\begin{figure}[ht!]
   \gridline{
   \fig{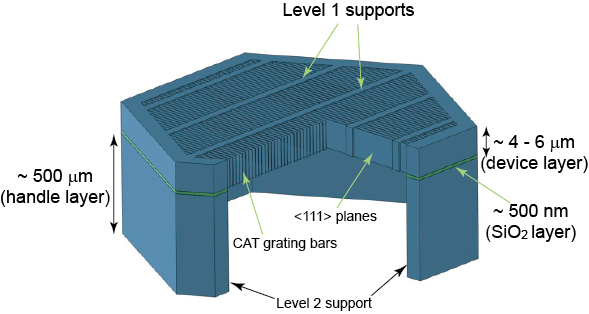}{0.43\textwidth}{(a)}
   \fig{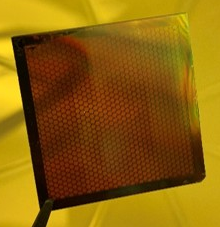}{0.2\textwidth}{(b)}
   \fig{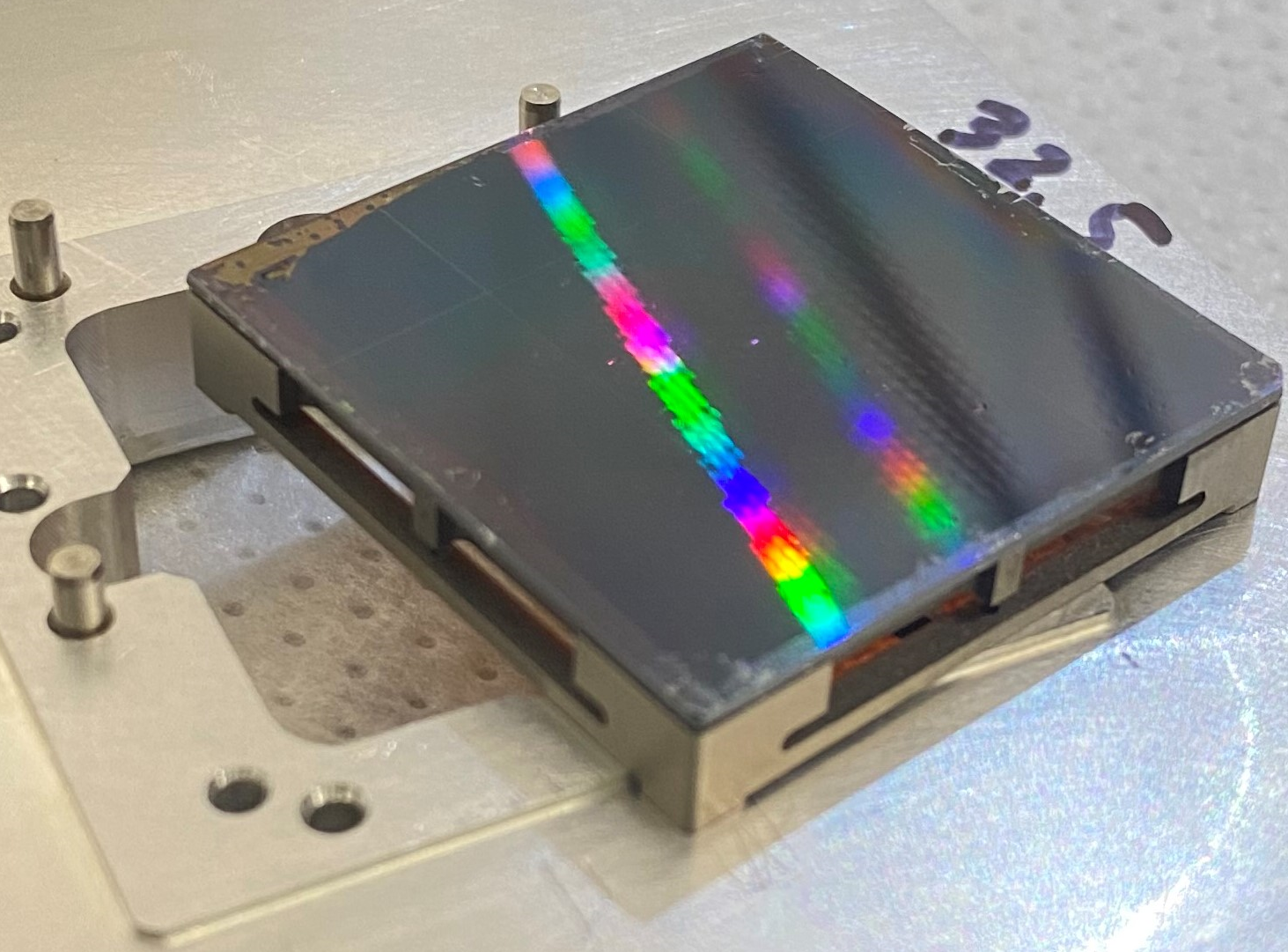}{0.27\textwidth}{(c)}
   }
\caption{Structural hierarchy of CAT gratings. (a) Schematic (not to scale) showing CAT grating bars, L1 cross supports, front-back-side aligned hexagonal L2 supports, and the buried SiO$_2$ layer separating front and back sides.  (b) Photograph of a silicon grating membrane. Only the hexagonal L2 mesh and the 2 mm-wide L3 frame can be seen by eye.  (c) Photograph of a grating facet, consisting of a silicon membrane bonded to a Ti flexure frame.  Visible diffraction is caused by the L1 mesh. \label{fig:hier}}
\end{figure}

Each Arcus OC requires $> 1000$ cm$^2$ of grating area, which is achieved by tiling reasonably-sized ($32\times32.5$ mm$^2$) grating facets.  Each facet consists of a patterned silicon membrane, bonded to a flexured titanium frame.  The membranes are fabricated from oxide-coated 200 mm silicon-on-insulator (SOI) wafers.  The $\sim 0.6$ mm-thick SOI wafer handle layer (back side) is patterned with an array of 2 mm-wide frames  (level 3 or L3 frames) that match the Ti frame dimensions (see Fig.~\ref{fig:hier}).  Inside each L3 support a $\sim 1$ mm-pitch, high throughput hexagonal mesh (L2 support) is etched into the handle layer.  On the front side of the wafer is the [110] Si device layer (thickness $d$), with a buried oxide (BOX) layer between front and back side.  Three structures are monolithically etched into the device layer: The 200 nm-period CAT grating bars (parallel to one set of device layer \{111\} planes), an L1 linear cross-support mesh (5 $\mu$m period) perpendicular to the CAT grating bars, and a hexagonal L2 mesh that is aligned with the back side L2 mesh. This hierarchy of support structures keeps the CAT grating bars in place and forms a mechanically strong, perforated membrane-like structure with a large open-area fraction for soft x-ray transmission.

The front side oxide mask patterning is performed at MIT Lincoln Laboratory, using a commercial electron beam-written photomask
and 4X optical projection lithography (OPL) (\citep{SPIE2020}).  The photomask contains the CAT grating, L1 and L2 patterns.
The CAT grating bars are patterned parallel to the \{111\} planes of the [110] device layer.  The back side oxide is patterned with an L2 mesh (aligned to the front side L2) and the L3 frames with a maskless aligner tool at the MIT.nano nanofabrication facility.  The front side oxide mask pattern is etched through the device layer using an off-campus deep reactive-ion etching (DRIE) tool, stopping on the BOX layer.  The wafer is then broken into chips containing one L3 frame each.  A short wet etch in potassium hydroxide solution smooths the CAT grating bar sidewalls until they largely consist of \{111\} planes (\citep{Alex_JVST2013}).  The front side is then protected, bonded to a carrier wafer, and the back side is DRIE'd, again stopping on the BOX.  The BOX layer is thinned via reactive-ion etching (RIE), and the chip is debonded, cleaned and critical-point dried.  Finally, the remaining BOX and oxide masks are removed with a hydrofluoric acid vapor etch, resulting in a Si membrane with freestanding CAT grating bars that are suspended between L1 cross supports (\citep{EIPBN2016}) (see Fig.~\ref{fig:CAT_schematic}(b)).

\section{grating alignment}

\begin{figure}[ht!]
    \gridline{
        \fig{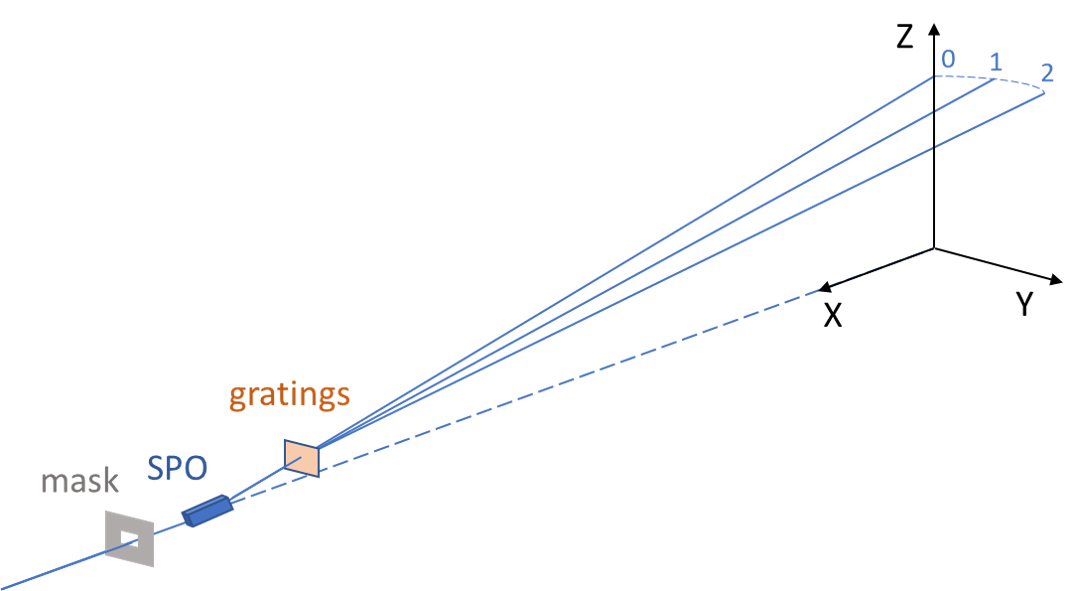}{0.63\textwidth}{(a)}
        \fig{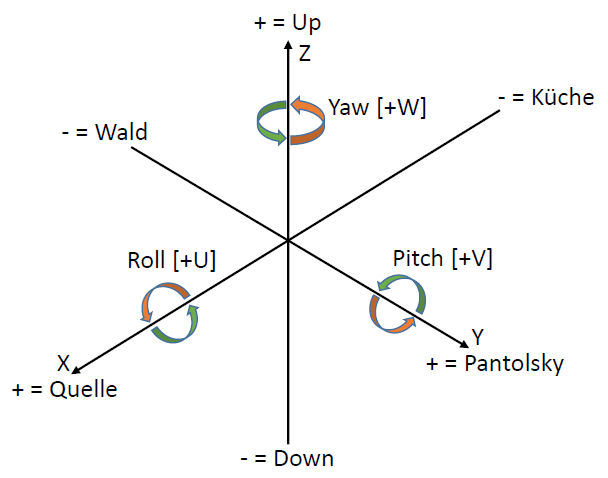}{0.37\textwidth}{(b)}
        }
\caption{(a) Schematic of the experimental setup at PANTER (not to scale).  X rays are diverging from the source ca.~120 m in $+x$ from the mask.  The mask defines the beam size incident onto the SPO, which focuses the x rays via double reflection to a point 13.258 m away, labeled ``0". The CAT gratings downstream of the SPO diffract x rays along the y-axis. Only focused diffraction orders $m = 1$ and 2 are shown.  See Section \ref{sec:exp} for details.  (b) Definition of rotation axes in the PANTER coordinate system. Quelle is the source side of the facility, K\"uche is the detector side.  Yaw is set for blazing in the (horizontal) +y direction.  When pitch and roll are zero, the CAT grating bars are parallel to the z-axis. \label{fig:rot}}
\end{figure}

Compared to reflection gratings, alignment tolerances for transmission gratings are significantly more relaxed in several degrees of freedom.  This is especially pronounced for x-ray diffraction where the diffraction angles of relevant orders are just a few degrees from the $0^{\mathrm {th}}$ order straight-through beam (\citep{SPIE2009}). The optics point-spread function (PSF) for each Arcus OC is anisotropic (\citep{Cash1991}), with an expected LSF around 3 arcsec full-width-half-max (FWHM) in the grating dispersion direction and less than 10 arcsec half-energy-width (HEW) in the cross-dispersion (CD) direction.  Nevertheless, even the tightest rotational grating-to-grating alignment requirements are only 15 arcmin ($3\sigma$) in grating roll (rotation around the grating normal, which rotates the dispersion axis) and 9 arcmin ($3\sigma$) in yaw (rotation around the grating bars, which affects the blaze angle), based on the current Arcus alignment error budget, which was updated based on the most recent optical alignment tests.  Some error components with weak effect on the overall system $R$ have been relaxed significantly since the 2021 Arcus proposal and previous work (\citep{moritz_SPIE2018}).  
Performance is even less sensitive to variations in pitch.  Fig.~\ref{fig:rot} shows a schematic of our experimental setup and the definition of the coordinate system and grating rotation axes.

Grating pitch and yaw can simply be measured by reflecting a visible light beam off the grating surface, but visible light reflected beams are insensitive to grating roll around the surface normal.  For 200 nm-period gratings only the $-1^{\mathrm {st}}$ (back-diffracted) order is accessible at large reflection angles with UV light.  Instead we use visible-light (HeNe laser) diffraction in transmission from the L1 cross-support mesh, which is lithographically defined in the same OPL mask for all gratings at 90 degrees from the CAT grating bar orientation (\citep{SPIE2021}).

Grating yaw adjustment is complicated by the fact that DRIE generally produces grating bars that are not perfectly normal to the device layer surface.  This so-called bar tilt (typically up to a few tenths of a degree) leads to a shifting of the blaze peak in angle, which can shift the intensity distribution towards diffraction orders that do not fall onto the readout.
We address this by measuring the bar tilt using a combination of small angle x-ray scattering (SAXS) and laser reflection (\citep{JungkiEIPBN}) and correct for it when bonding the silicon membrane to the facet frame.

For alignment and bonding we use an updated version of our Grating Facet Assembly Station (GFAS) (\citep{SPIE2018}). Briefly, a precision hexapod holds the grating membrane, while grating pitch, roll, and yaw are measured.  The hexapod then rotates the membrane into the desired orientation, including the yaw correction angle derived from the bar tilt measurement.  An epoxy drop is deposited at the center of each L3 edge, and a high-repeatability vertical translation stage lowers a facet frame to a fixed height (nominally 0.2 mm from the membrane), compressing all four epoxy drops without making direct contact with the membrane.  Four fibers provide a UV flash to cure the epoxy.  The bar tilt correction is now fixed via the different thicknesses of the four cured epoxy bonds.  After removal from the GFAS the grating facet undergoes a thermal cure.  The goal of this procedure is to produce interchangeable grating facets that can be aligned and mounted to larger mechanical structures (so-called grating windows), using the metal facet frames as mechanical alignment references. 

For this work four grating facets were produced.  Two of them were fabricated using membranes from two separate 200 mm SOI wafers with nominally 4 $\mu$m-thick device layers (CNS1 and CNS5), and two using membranes from a single SOI wafer with nominally 6 $\mu$m-thick device layers (SEG25 and SEG30).  The front sides of the latter two were patterned with a previous OPL mask that only contained the L1 and CAT grating features, but no front side L2 structure.

Alignment and bonding of membranes to frames was done according to our current procedure.  Unfortunately, due to schedule constraints we could not obtain a grating window plate in time that met all of its specifications, and we did not have time to perform proper facet-to-window alignment for the four facets, which mainly affects grating roll.  

\section{Experimental setup}
\label{sec:exp}

\begin{figure}[ht!]
   \begin{center}
   \begin{tabular}{ c c } 
   \includegraphics[height=6cm]{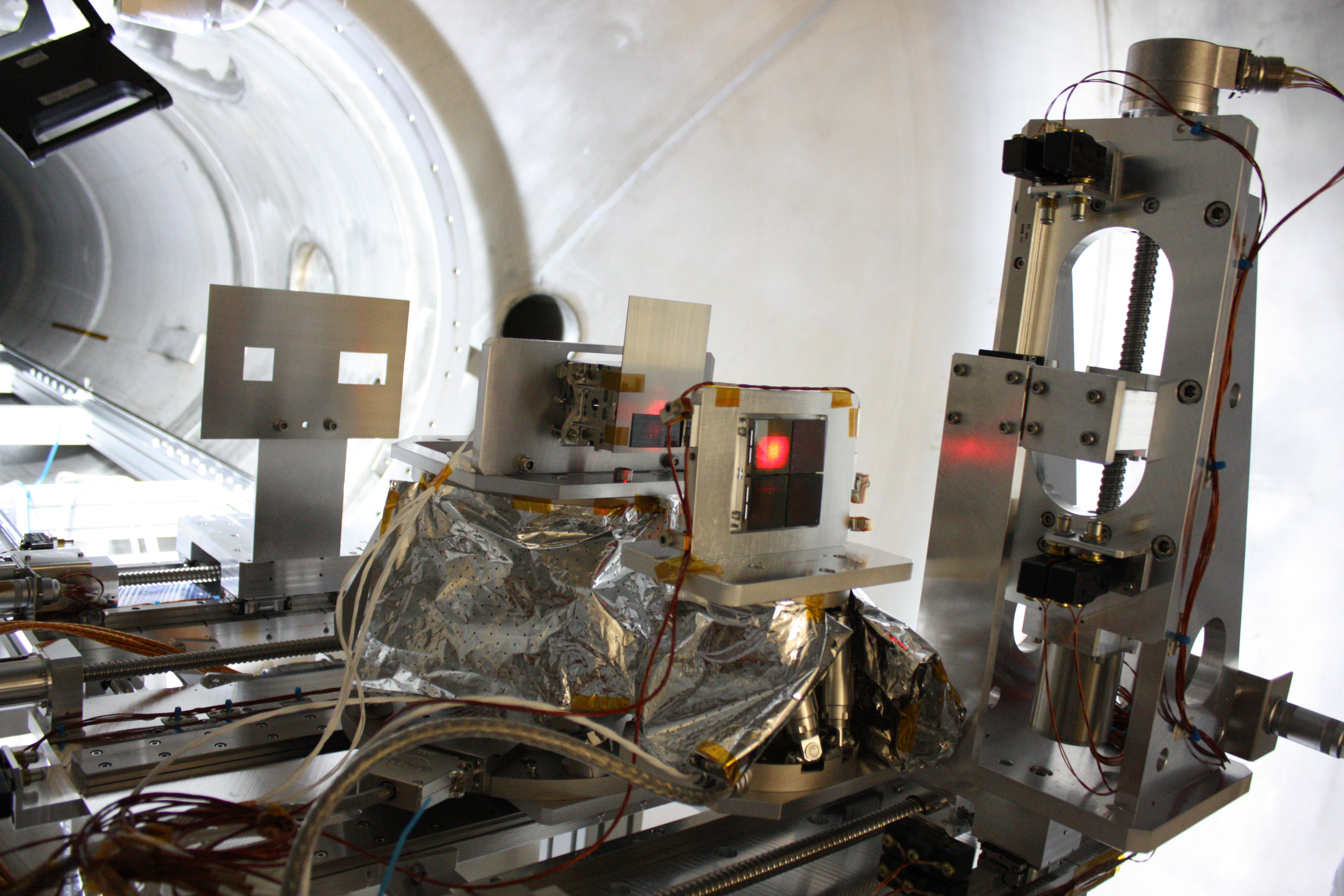}
   \includegraphics[height=6cm]{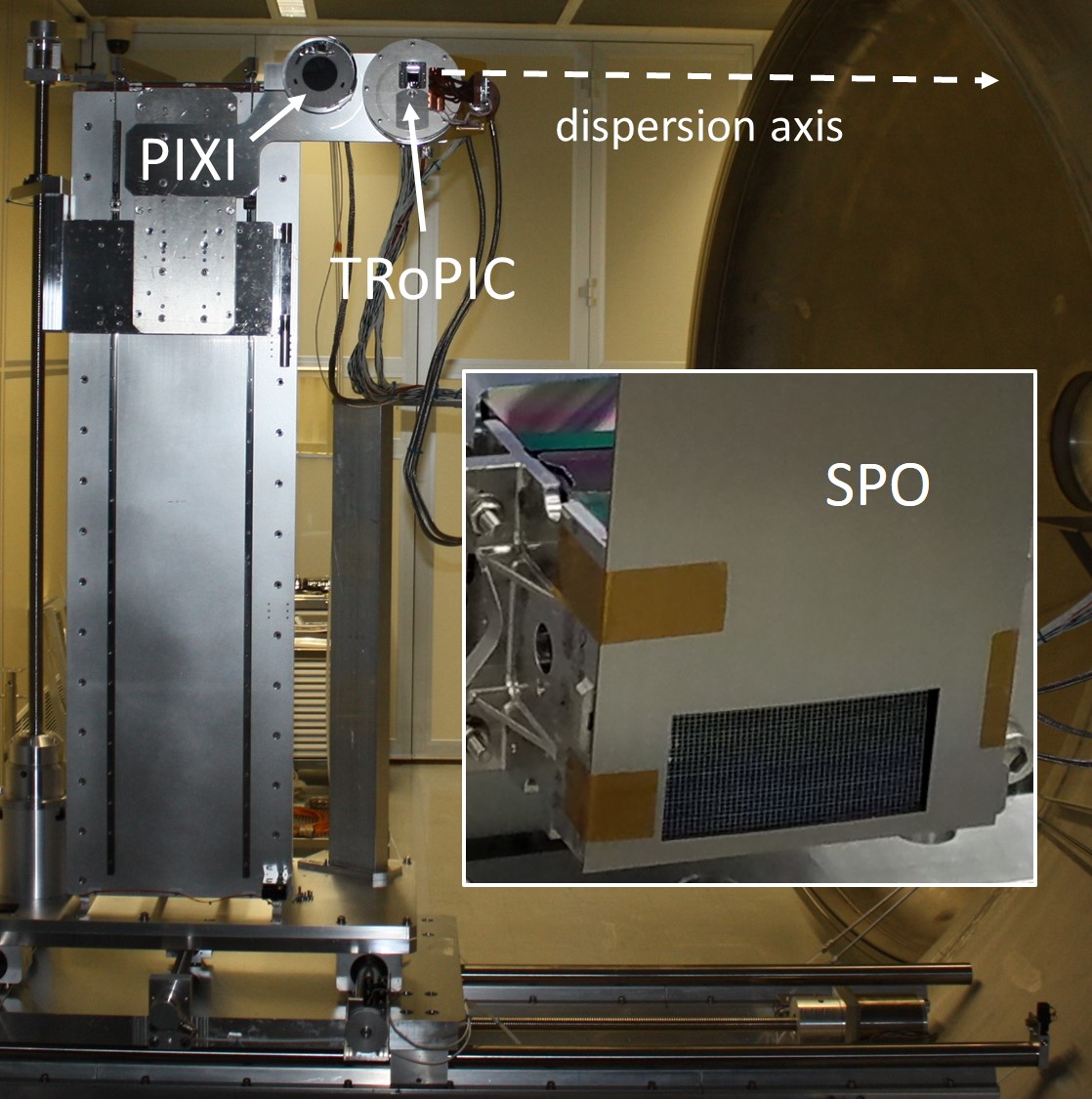}
   \end{tabular}
   \end{center}
\caption{Left: Picture of the setup at PANTER before sliding the carriage into the beam tube.  The x-ray source is far outside of the picture to the left.  From left to right are the aperture plate, the masked-off SPO, the grating window on top of its hexapod, and the grating window vertical translation stage.  The alignment laser is exiting the bottom XOU and transmitted though grating CNS1 (top left of the grating window).  Right: Downstream view toward the cameras mounted to the top of the camera xyz translation stage stack. Insert: Close-up view of the masked-off SPO, with only the bottom XOU used for grating illumination. \label{fig:setup}}
\end{figure}

The grating window was taken to the PANTER x-ray facility of the Max-Planck-Institut f\"ur Extraterrestrische Physik.  It features an electron-bombardment x-ray source on one end, connected via a 120 m-long vacuum pipe to a 12 m-long, 3.5 m-diameter vacuum chamber on the other end.  A carriage in the chamber held one hexapod on a horizontal translation stage to align SPO mirror module MM-0036 (see Fig.~\ref{fig:setup}).  Only the bottom X-ray Optical Unit (XOU-0078, outermost radius of curvature = 737 mm) was used and masked off to an aperture of $\sim 46\times 22.3$ mm$^2$ (azimuth $\times$ radius).  A second hexapod holding the grating window was mounted with the gratings 136 mm downstream of the SPO node on a $y$-$z$ (horizontal-vertical) translation stage stack.  Upstream of the SPO was another horizontal translation stage with an aperture plate with two apertures: One slightly larger than the XOU mask ($50 \times 30$ mm), and a small mask of size $\sim 28 \times 30$ mm$^2$ for illuminating a single grating.  At the far end of the chamber is an $xyz$-stack of translation stages for two side-by-side detectors: A Princeton Instruments (PIXI) x-ray integrating CCD camera ($1300 \times 1340$ 20 $\mu$m pixels), and the energy-sensitive TRoPIC CCD camera ($256 \times 256$ 75 $\mu$m pixels).

A red laser originating from the direction of the x-ray source was used for preliminary alignment by eye.  First the SPO was aligned for proper double reflection of the laser beam in the vertical direction ($\sim 3.5$ deg from horizontal), then the back-reflection from a grating onto the SPO was used to align the grating to normal incidence from the beam exiting the optic.  Grating roll was checked for all four gratings by recording the roll angle for which the $1^{\mathrm {st}}$ order diffraction from the L1 mesh was vertically below the $0^{\mathrm {th}}$ order at the detector plane.

Next the carriage was moved upstream into the beam pipe to bring the finite-source-distance focus of the optic near the center of the camera x-stage range and alignment was re-checked using the source laser. Grating roll was measured again. Finally, the chamber was closed and evacuated.

The laser-based roll measurements, performed by eye and with measuring tape, showed that gratings SEG25 and SEG30 only differed by $\Delta U = U_{SEG25} - U_{SEG30} = -1.5$ arcmin, with an estimated measurement uncertainty of 5 arcmin.  Gratings CNS1 and CNS2 had $\Delta U$ of 33 arcmin and -52 arcmin, respectively.  See Fig.~\ref{fig:rot}(b) for the definition of $U$.

\section{Measurement Results}

\begin{figure}[ht!]
   \begin{center}
   \begin{tabular}{ c c } 
   \includegraphics[height=4cm]{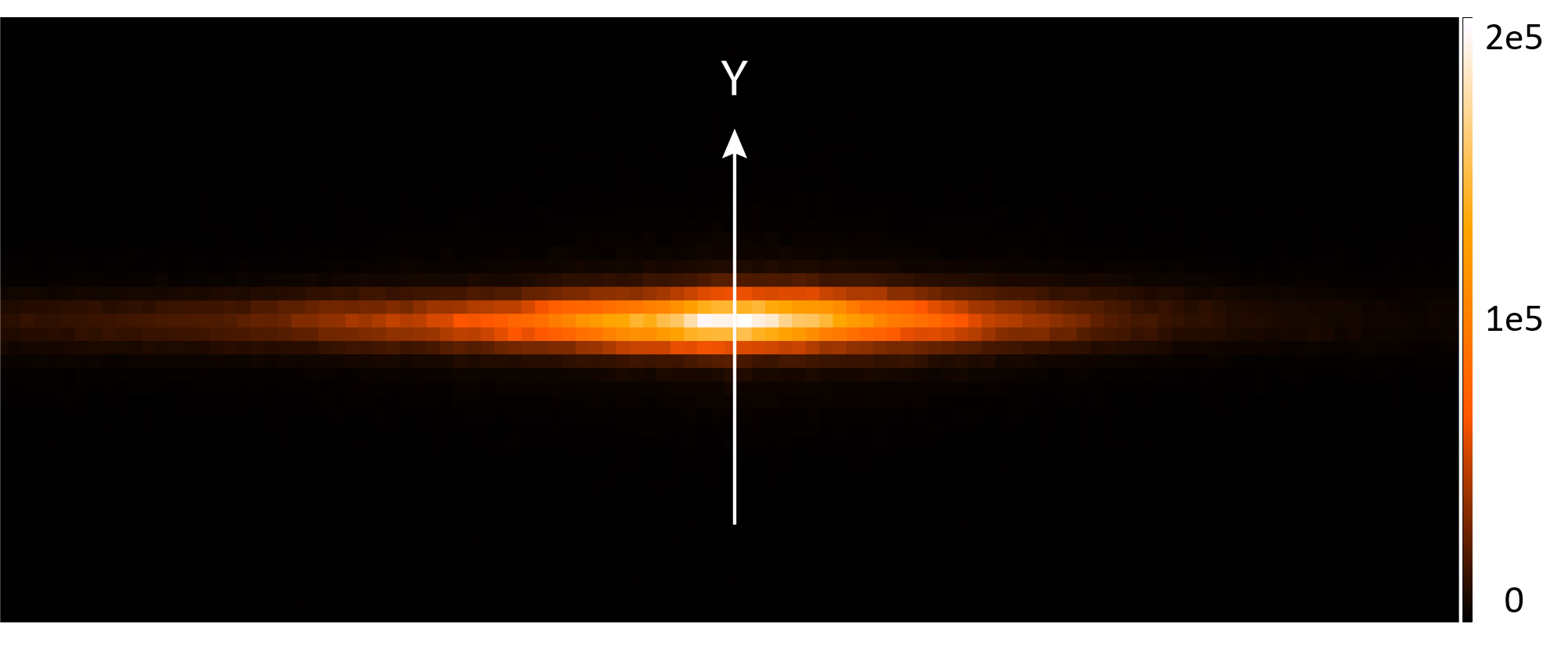}
   \includegraphics[height=4cm]{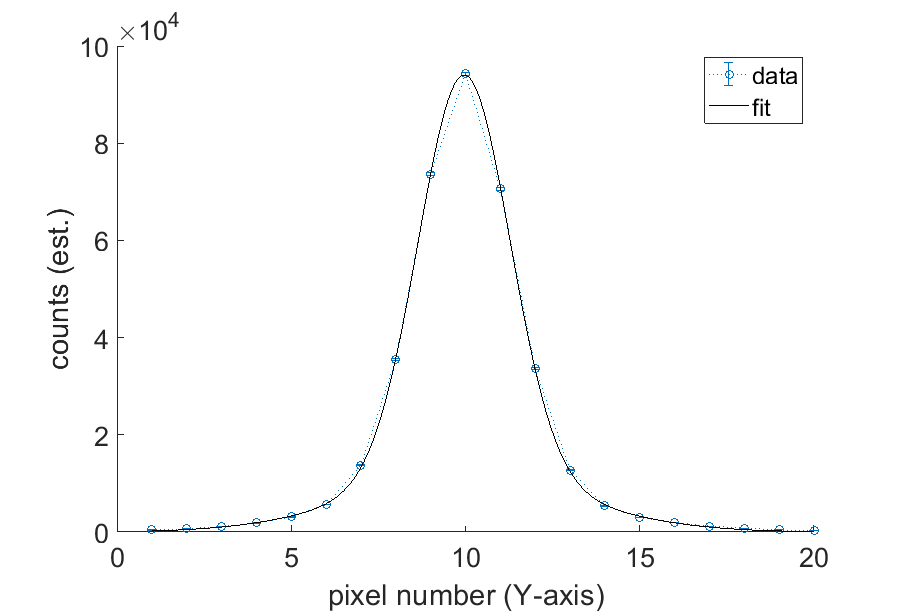}   
   \end{tabular}
   \end{center}
\caption{Best focus direct beam.  Left: Image of the direct beam with the small mask taken with the PIXI camera (linear intensity scale). The individual 20 $\mu$m pixels can be discerned.  The arrow indicates the +y direction, which is also chosen as the CAT grating dispersion direction.  The PSF anisotropy is typical for reflection at small angles of grazing incidence with azimuthal sub-aperturing (\citep{Cash1991,SPIE2010}).  Right: Projection onto the dispersion axis and fit to a double Gaussian.  The FWHM is $0.944 \pm 0.020$ arcsec in $y$ and $\sim 8.9$ arcsec in $z$ (HEW $\sim 9.6$ arcsec).  The plate scale is 0.31 arcsec/pixel.
\label{fig:DB}}
\end{figure}

After the required vacuum level was reached and the Al anode selected, the alignment of all components relative to the source and relative to each other was fine-tuned using x rays.  With the large mask, but without gratings in the beam, the best focus of the optic was found with a camera x-scan.  We then performed a horizontal scan of the small mask behind the optic to find the position with the smallest FWHM in the dispersion direction and repeated the focus scan.  Best focus was found 13258 mm from the optic node (see Fig.~\ref{fig:DB}).  A simple Gaussian fit to the beam profile projected on the dispersion (Y) axis gives a FWHM of 1.04 arcsec, but underestimates the tails of the beam.  A double Gaussian fits the data well, where the second, wider Gaussian is about eight times weaker and 2.5 times wider than the main, narrow one, which has a FWHM of 0.944 arcsec.
\label{meas}

\subsection{Relative yaw alignment between gratings}

\begin{figure}[ht!]
   \begin{center}
   \begin{tabular}{ c c } 
   \includegraphics[height=6cm]{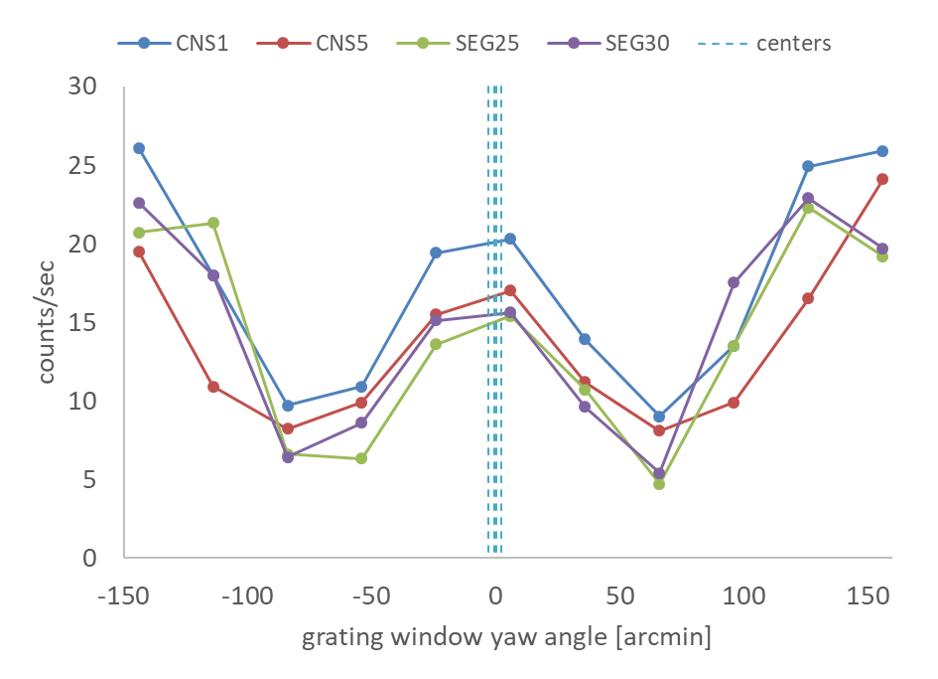}
   \end{tabular}
   \end{center}
\caption{TRoPIC count rate of $0^{\mathrm {th}}$ order for all four gratings as a function of relative hexapod yaw angle. The CCD energy selection was centered on the Al-K line.  Transmission in the $0^{\mathrm {th}}$ order peaks when the grating bars are parallel to the incident x rays.  As the bars are rotated, more x rays are blazed into higher orders, reducing $0^{\mathrm {th}}$ order flux.  At even larger angles the blazing effect is reduced again due to geometry, exceeding the critical angle for sidewall reflection, and due to significant transmission through the Si bars for Al-K x rays (E = 1.49 keV), resulting in increased $0^{\mathrm {th}}$ order transmission. \label{fig:yawscan}}
\end{figure}

The relative orientation of the CAT grating bars between the four gratings can be determined using x rays.  Deep-etched CAT grating bars have cross-sectional profiles that are fairly mirror-symmetric around an axis going through the center of the bar which defines bar tilt (\citep{JungkiEIPBN}).  We performed yaw scans for each grating, where the grating is centered on the SPO exit aperture defined by the small mask and rotated in the yaw direction.  As simulations predict, the measured diffraction efficiency in $0^{\mathrm {th}}$ order is symmetric around the yaw angle where the grating bars are on average parallel to the incident x rays.  As shown in Fig.~\ref{fig:yawscan}, comparing the yaw scan centers of symmetry for the four gratings we find relative yaw angles of -10, -12, -13, and -15 arcmin, resulting in a standard deviation of 1.8 arcmin from the average.  Due to time constraints only a small number of data points was taken for each grating, and integration was stopped around 500 counts for each point. We estimate an uncertainty of 0.5 arcmin ($1\sigma$) for the center of symmetry fits to these widely spaced measurement points. The relative yaw alignment between the four gratings meets the currently budgeted tolerance of 9 arcmin ($3\sigma$).

\subsection{Measurement of effective resolving power in $18^{\mathrm {th}}$ order Al-K$_{\alpha}$}

The measured line shape of a narrow emission line is determined by the shape of the line itself (e.g., its natural linewidth) the resolving power of the spectrometer.  In our case the spectrometer consists of a source of finite size, a focusing optic with a finite LSF, gratings, and detectors.  All four components, and their relative placement, impact the resolving power.  The goal of our measurements is to derive a lower limit for the minimum resolving power that can be achieved with the fabricated gratings.  We followed a similar but simpler approach than described previously in more detail (\citep{AO2019}).  The Al-K$\alpha_{1,2}$ doublet is well-modeled by the superposition of two Lorentzians with 2:1 intensity ratio.  First we measure the direct-beam LSF (LSF$_{DB}$) produced by the combination of source, SPO, and camera at best focus, and approximate it by a Gaussian of width $\sigma_{DB}$.  Then we move to the highest diffraction order we can reach in our setup where the dispersed width of the Al-K$_{\alpha}$ doublet is large compared to $\sigma_{DB}$ and measure the line shape of the doublet.  We then fit the data by convolving the dispersed natural line shape with a Gaussian of variable width  $\sigma_f$  and compare it to $\sigma_{DB}$.  Fitting parameters are an overall amplitude, position of the Al-K$_{\alpha_1}$ line, and $\sigma_f$.  The distance between the two K$_{\alpha}$ peaks and the Lorentzian widths on the detector are calculated based on the geometry of the setup and the doublet parameters from \citep{AO2019}.  The background is fit to a sloped straight line, giving a total of five fit parameters.  We generally (but not always) find $\sigma_f > \sigma_{DB}$ and conservatively attribute this additional broadening fully to grating imperfections such as period variations, and assign it a Gaussian width

\begin{equation}
    \sigma_G = \sqrt{\sigma_f^2 - \sigma_{DB}^2} \ . 
    \label{R}
\end{equation}
 
Based on Eq.~\ref{ge}, $\sigma_G$ can be converted into period variations $\Delta p$. Assuming a Gaussian period distribution with FWHM $\Delta p$ we define $R_G = p/\Delta p$ as the effective grating resolving power and derive a lower limit for $R_G$ from the fit to the data.

We optimized the source settings (filament heating voltage, Wehnelt cylinder voltage, acceleration voltage) to minimize the beam size at best focus and achieved a FWHM of $0.944\pm 0.020$ $(3\sigma)$ arcsec in the dispersion direction.

In the current setup we were able to reach $18^{\mathrm {th}}$ order for 0.834 nm wavelength (Al-K$_{\alpha}$) with the PIXI camera at an angle of $(\theta + \beta_m) \approx 4.3$ degrees  (almost 1 m) from $0^{\mathrm {th}}$ order.   
Each grating was measured individually, nominally rotated by $\sim 2.15$ degrees in yaw toward the $18^{\mathrm {th}}$ order for maximum blazing.  While this greatly improves efficiency in $18^{\mathrm {th}}$ order compared to normal incidence, this angle is much larger than the critical angle between a Si surface and Al-K x rays ($\theta_c \sim 1.2$ degrees).  As a result, the count rates were low, requiring long exposure times.  We determined the best focus position for $18^{\mathrm {th}}$ order via ray trace.  
Grating roll for each grating was adjusted such that the camera did not have to be moved and $18^{\mathrm {th}}$ order fell on the same vertical range of detector pixels within $\approx \pm 75$ $\mu$m.  No roll adjustment was made between gratings SEG25 and SEG30.

The PIXI detector does not count individual photons, but is an integrating detector, reporting analog-to-digital units (ADUs).  Based on previous work we estimate that a single Al-K$_{\alpha}$ photon, when detected in a single pixel without pile-up, results in a signal of circa 80 ADUs.  We then simply divide the number of ADUs measured in each pixel by 80 to obtain an estimate of the number of Al-K$_{\alpha}$ photons.  Summing along each detector column (CD direction) and assuming Poisson statistics for the estimated photon count rates in each column we perform least-square fitting.  To estimate the uncertainty in the best fit results for $R_G$ we repeat fitting for fixed higher and lower values of $R_G$ until $\chi^2$ has increased by $\Delta\chi^2 = 9$ (99.73\% or $3\sigma$ uncertainty).

\begin{figure}[ht!]
   \begin{center}
   \begin{tabular}{ c c } 
    \includegraphics[height=6cm]{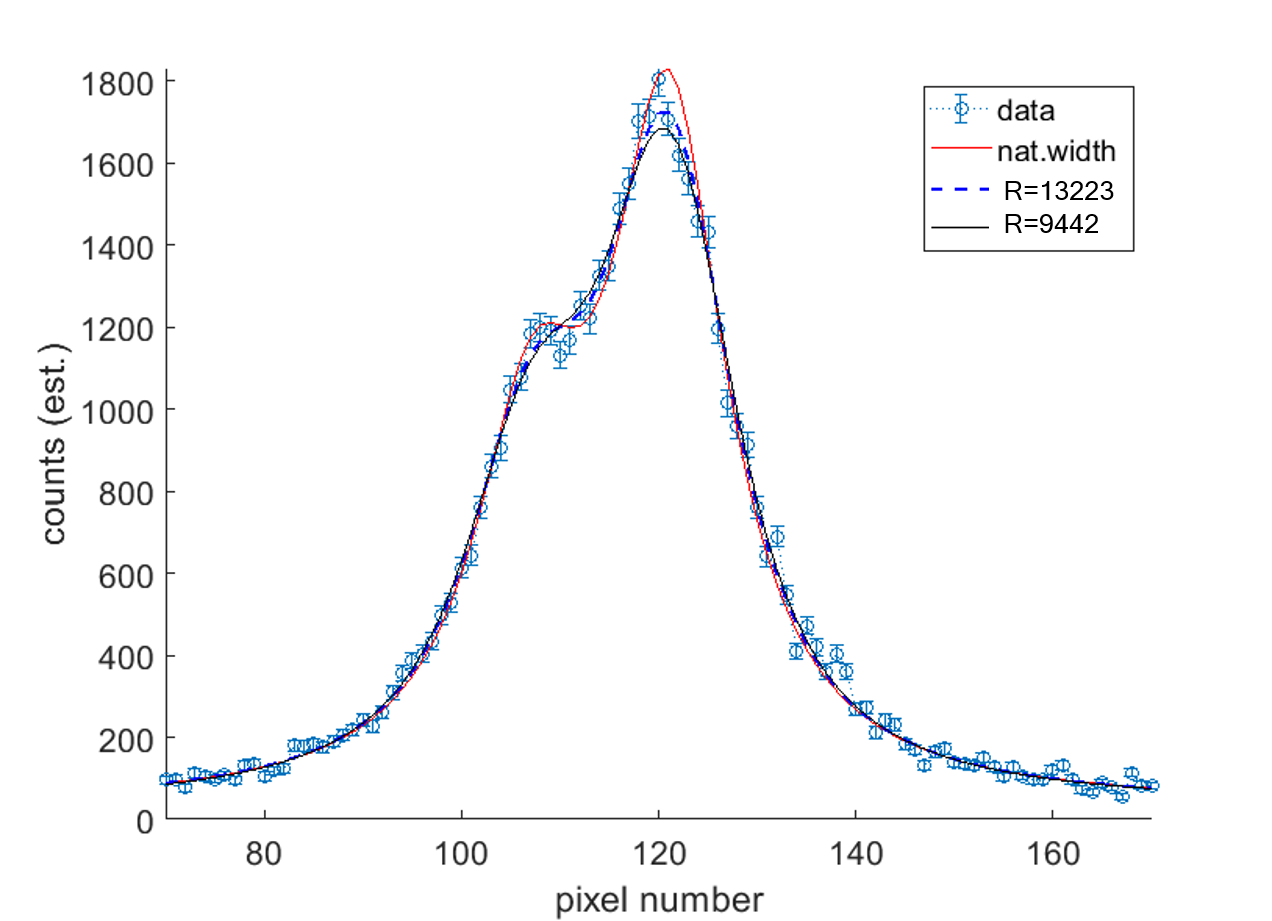}
    \includegraphics[height=6cm]{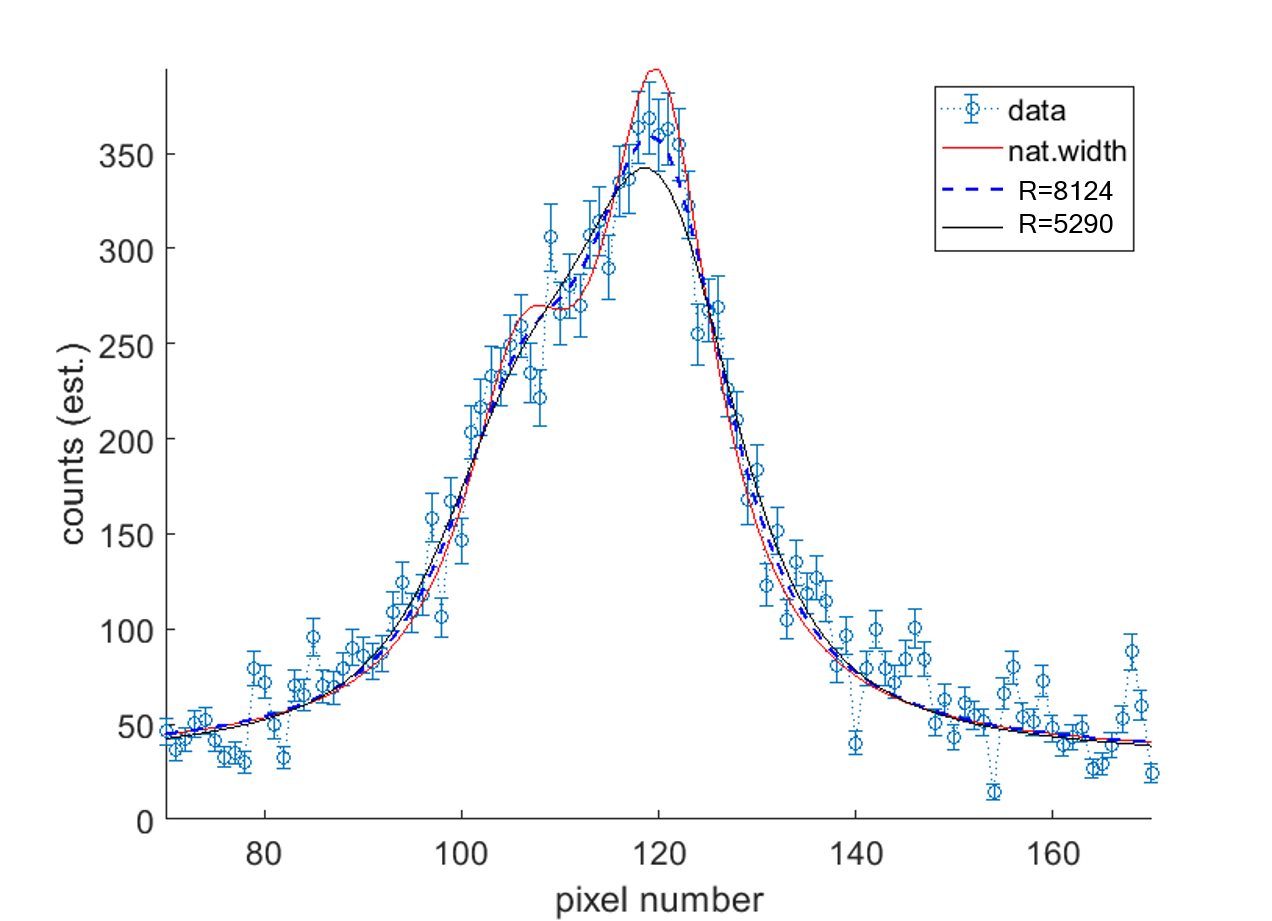}
   \end{tabular}
   \end{center}
\caption{Al-K$_{\alpha}$ doublet in $18^{\mathrm {th}}$ order.  Shown are the estimated number of Al-K photons as a function of detector pixel column number.  The red line is the natural line shape of the doublet, the dashed line is the best fit to the data.  The difference in counts is primarily due to the different integration times, but also to differing diffraction efficiencies.  Left: Grating SEG25.  The black solid line is the curve for the lower $3\sigma $  confidence limit, corresponding to $R_G = 9442$.  Right: Same for grating CNS1.  The solid line corresponds to the lower limit of $R_G = 5290$.\label{fig:18th-2}}
\end{figure}

All four gratings showed high effective resolving power with best fit values between $R_G \sim 6900$ and $\sim 13000$ (see Fig.~\ref{fig:18th-2} for examples). The 6 $\mu$m-deep gratings have higher count rates and higher $R_G$.  The key indicator of finite $R_G$ is the slightly more rounded shoulder on the left of the doublet, and the less-pronounced main peak compared to the natural line shape.  Lower counting statistics lead to larger uncertainties and smaller lower confidence limits for $R_G$.

\begin{figure}[ht!]
   \begin{center}
   \begin{tabular}{ c c } 
   \includegraphics[height=6cm]{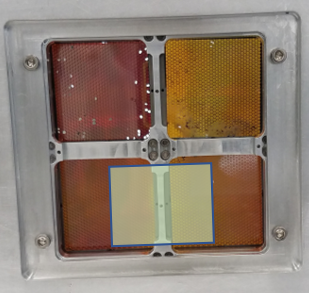}
    \includegraphics[height=6cm]{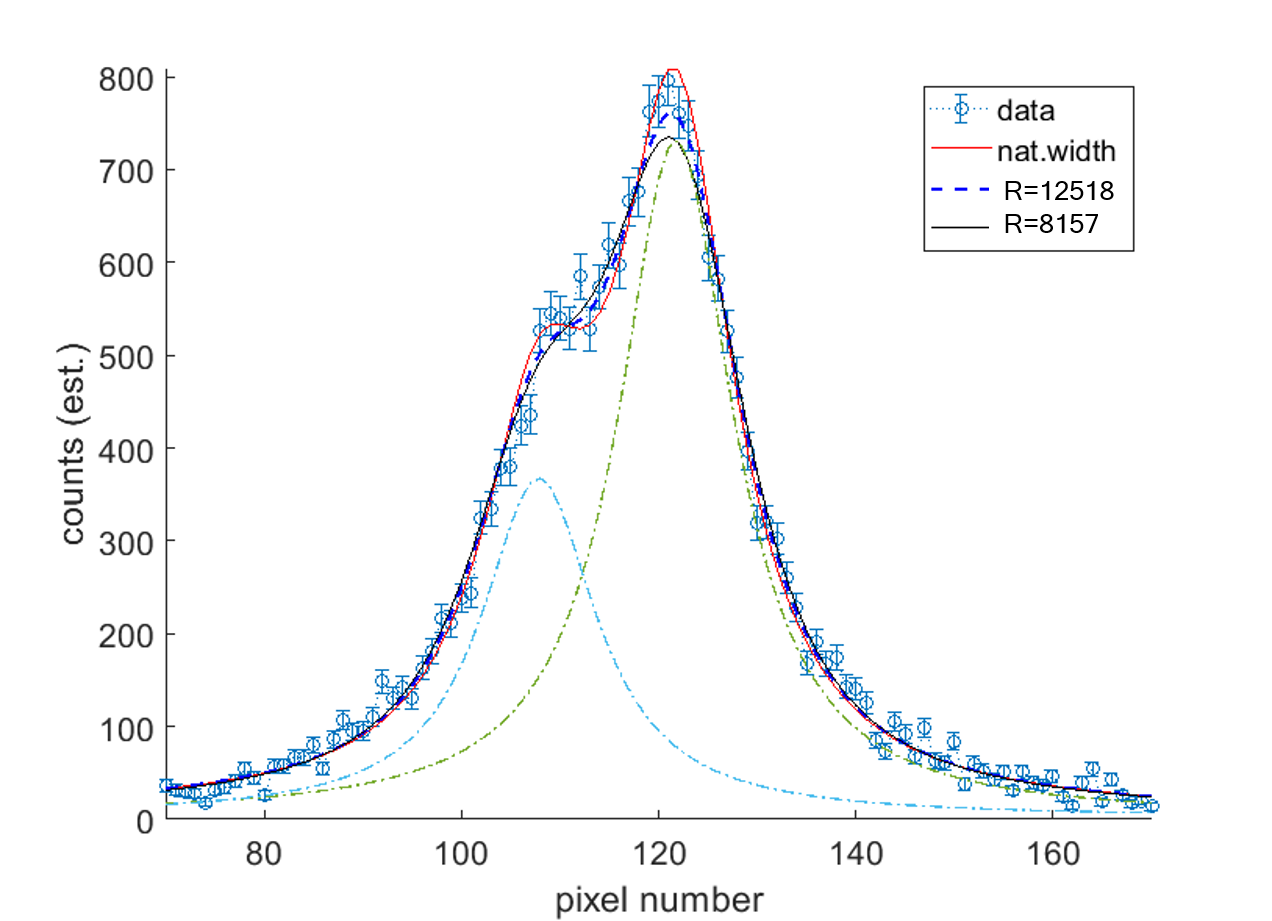}
   \end{tabular}
   \end{center}
\caption{Simultaneous illumination of gratings SEG25 and SEG30.  Left: Photograph of the grating window with an overlayed rectangle approximating the area of illumination.  (The upper left grating (CNS5) suffered some blown-out hexagon interiors due to excessive agitation during the piranha etch cleaning step after debonding from the carrier wafer.)  Right: Al-K$_{\alpha}$ doublet spectrum measured in $18^{\mathrm {th}}$ order.  The red line is the natural width of the doublet, the dashed line is the best fit to the data, and the black solid line is the curve for the lower $3\sigma $ confidence limit, corresponding to $R_G = 8157$. The dash-dotted lines show the individual K$\alpha_1$ and K$\alpha_2$ components with their natural widths on top of the weak sloped background. \label{fig:18th}}
\end{figure}

We also measured $18^{\mathrm {th}}$ order from the simultaneous partial illumination of gratings SEG25 and SEG30.  To avoid bias, we first found a common yaw angle where both gratings gave the same count rate.  Then we moved the midpoint between the two gratings to the center of the small aperture (see Fig.~\ref{fig:18th}). The effective resolving power from the combination of the two gratings was found to be $R_G \sim 1.3\times 10^4$.  

Results for $R_G$ for all four gratings and $3\sigma$ uncertainties are listed in Table \ref{tab:R}.  The lowest lower bound is still near $R_G = 5000$, and the data for the combined illumination of two gratings is compatible with $R_G = \infty$ with $> 99.73$\% confidence. 

\begin{table}
\begin{center}
\caption{Fit results for effective resolving power $R_G$ for all four gratings and for the partial combination of SEG25 and SEG30. \label{tab:R} }
\begin{tabular}{l|c|c|ccc}
Grating & diffraction & total & best fit & upper bound & lower bound  \\
 & order & counts* & $R_G$ & ($3\sigma$) & ($3\sigma$) \\
\hline
CNS1$^a$ & 18 & 14401 & $8.1 \times 10^3$ & $2.6 \times 10^4 $ & $5.3 \times 10^3$ \\
CNS5$^a$ & 18 & 11315 & $6.9\times 10^3$ & $1.3\times 10^4$ & $4.9\times 10^3$  \\ 
SEG25$^b$ & 18 & 54151 & $1.3\times 10^4$ & $3.2\times 10^4$ & $9.4\times 10^3$ \\
SEG30$^b$ & 18 & 163401 & $9.3\times 10^3$ & $1.1\times 10^4$ & $8.1\times 10^3$ \\
SEG25/30 & 18 & 23287 & $1.3\times 10^4$ & $\infty$ & $8.2\times 10^3$ \\
\hline
CNS1 & 21 & 4132 & $7.1\times 10^3$ & $\infty$ & $3.7\times 10^3$ \\
CNS5 & 21 & 3498 & $7.0\times 10^3$ & $\infty$ & $4.1\times 10^3$  \\ 
SEG25 & 21 & 4169 & $1.1\times 10^4$ & $\infty$ & $5.1\times 10^3$ \\
SEG30 & 21 & 5443 & $9.2\times 10^3$ & $\infty$ & $5.2\times 10^3$ \\
SEG25/30 & 21 & 2061 & $\infty $ & $\infty$ & $5.1\times 10^3$ \\
\hline
 \end{tabular}
\end{center}
{\small
 *Counts for $18^{\mathrm th}$ order are estimates.  $^a$ Nominally 4 $\mu$m-thick device layer. $^b$ Nominally 6 $\mu$m-thick device layer.
   }
\vspace{-0.1in}
\end{table}

\subsection{Measurement of effective resolving power in $21^{\mathrm {st}}$ order Al-K$_{\alpha}$}

The cameras used in this work were mounted side-by-side.  By way of serendipity the TRoPIC camera was positioned near the angle of diffraction for the $21^{\mathrm {st}}$ order during PIXI measurements of the $18^{\mathrm {th}}$ order.  While TRoPIC has much larger pixels (75 $\mu$m) than PIXI (20 $\mu$m), we can 
utilize the information contained in the charge ratios of events split over more than one pixel and improve the spatial resolution significantly to the sub-pixel scale.
We processed TRoPIC data for the $21^{\mathrm {st}}$ order using the resolution-enhancing algorithms described in \citep{Dennerl_SPIE2012}.  The data is then projected onto the dispersion axis and binned on a 20 micron grid.  We only took a brief exposure of the direct beam with TRoPIC.  It was processed in the same way as the $21^{\mathrm {st}}$ order data and resulted in a FWHM of $1.15 \pm 0.06$ $(3 \sigma)$ arcsec in the dispersion direction.  We used the corresponding value of $\sigma_{DB}$ in the calculation of $R_G$ for $21^{\mathrm {st}}$ order.

\begin{figure*}
\gridline{
          \fig{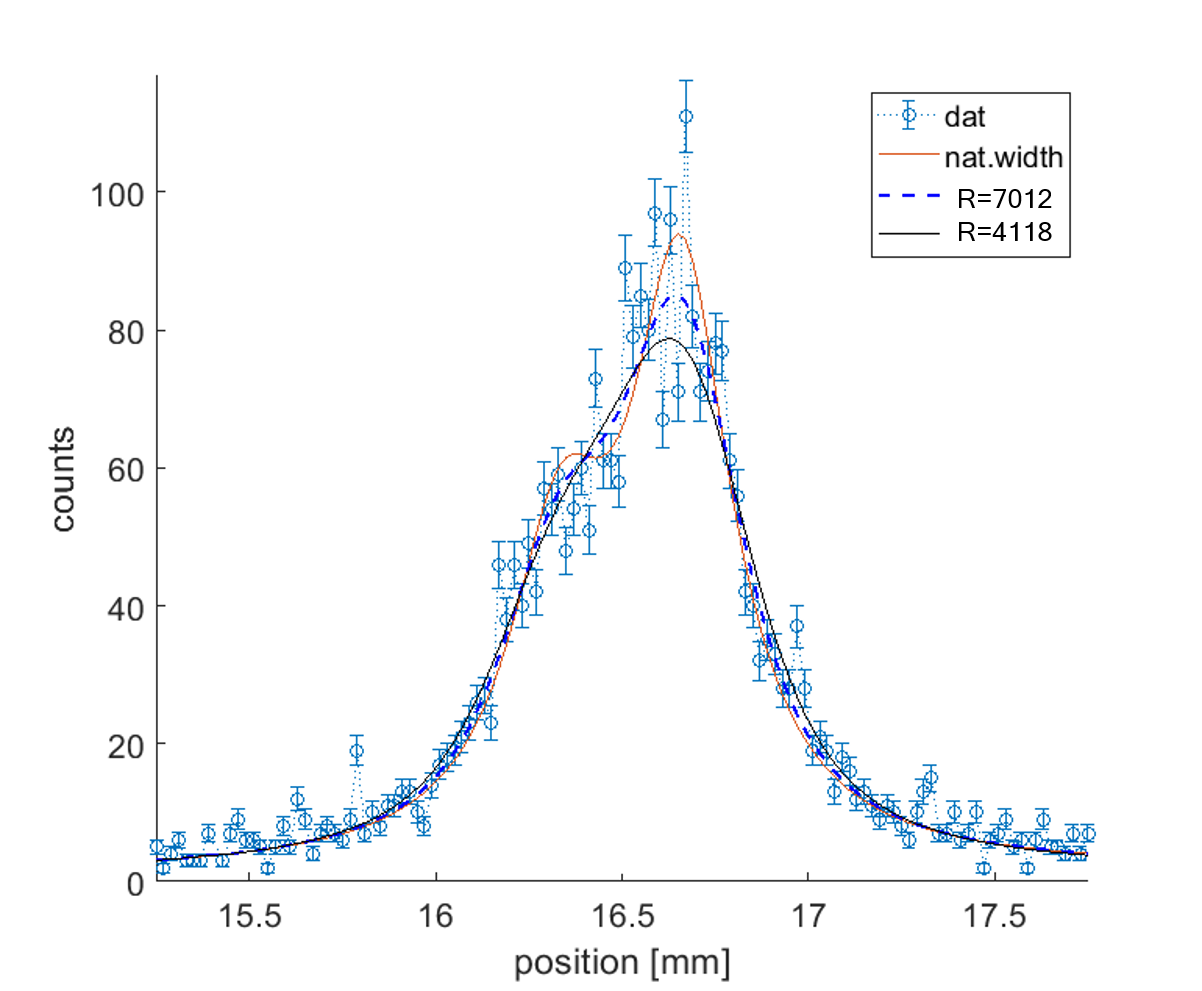}{0.5\textwidth}{(a)}
          \fig{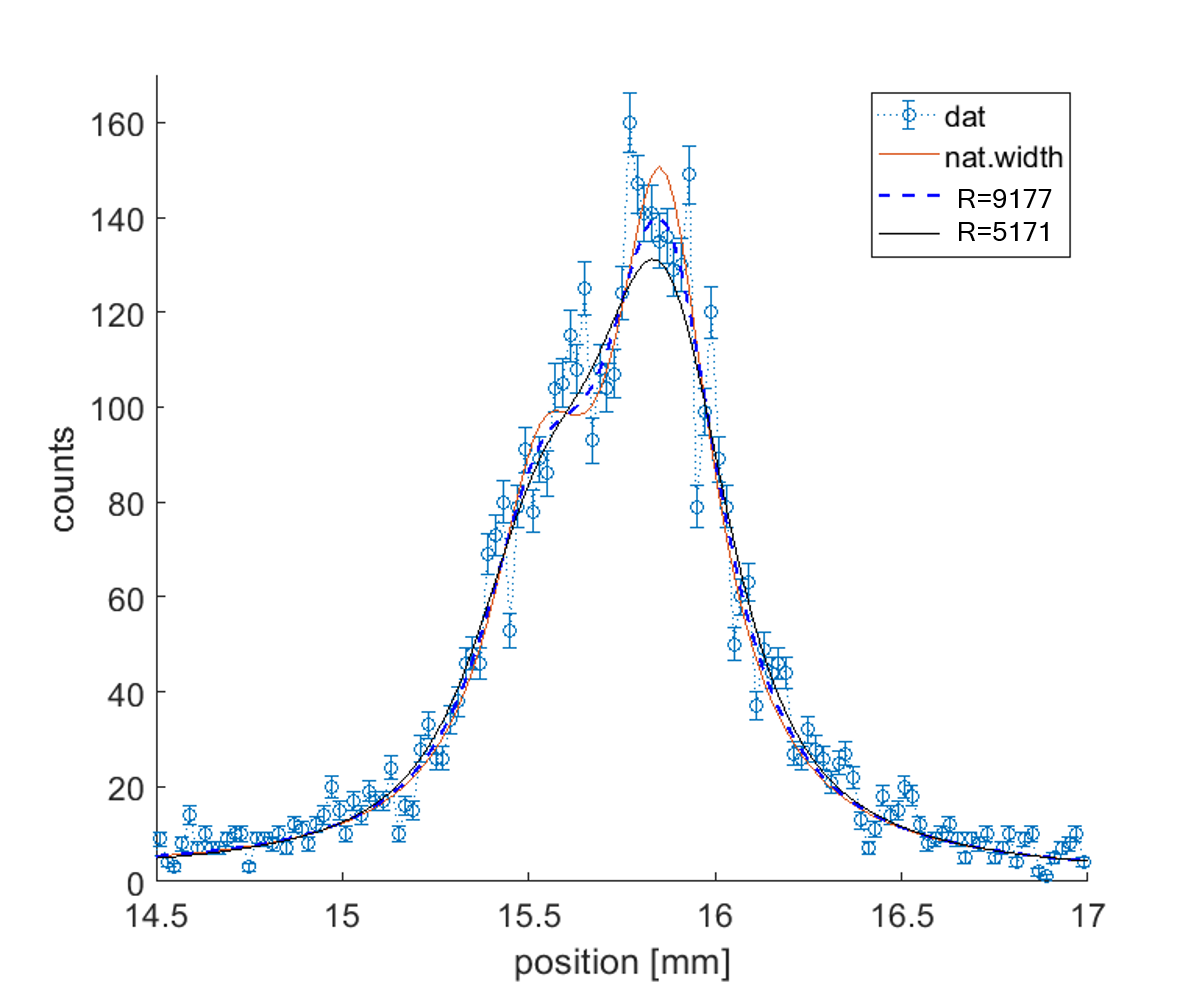}{0.5\textwidth}{(b)}
          }

\caption{Al-K$_{\alpha}$ doublet in $21^{\mathrm {st}}$ order.  Shown are the number of Al-K photons as a function of position on the detector for gratings CNS5 (a) and SEG30 (b) as examples.  The red line in each figure shows the natural width of the doublet, the dashed line is the best fit to the data, and the black solid line is the curve for the lower $3\sigma $ confidence limit.  (The doublets are in a different position on the detector because the data were taken at different $x$ or focus positions of the camera.)
\label{fig:21st}}
\end{figure*}

The $21^{\mathrm {st}}$ order provides much lower count rates, but TRoPIC offers the benefits of single photon counting and energy resolution, allowing us to reject most of the weak bremsstrahlung continuum.  The resulting data is noisier than the $18^{\mathrm {th}}$ order data and has wider $R_G$ uncertainties, but the best fit $R_G$ values, following the same fitting process as for the $18^{\mathrm {th}}$ order, are still in the 7000 to $\infty$ range.  Fig.~\ref{fig:21st} shows two examples of data, best fits, and $3 \sigma$ uncertainty curves.  Due to the lower counting statistics all data sets are compatible with $R_G = \infty$ well within the $3\sigma$ bounds.  For the combination of SEG25 and SEG30 the best fit gives $\sigma_f$ slightly smaller than $\sigma_{DB}$, implying no measurable broadening from the gratings. Detailed numbers and total counts are listed in Table \ref{tab:R}.

\subsection{Measurement of diffraction efficiency at O-K wavelength}

\begin{figure}[ht!]
   \begin{center}
   \begin{tabular}{ c c } 
   \includegraphics[height=6cm]{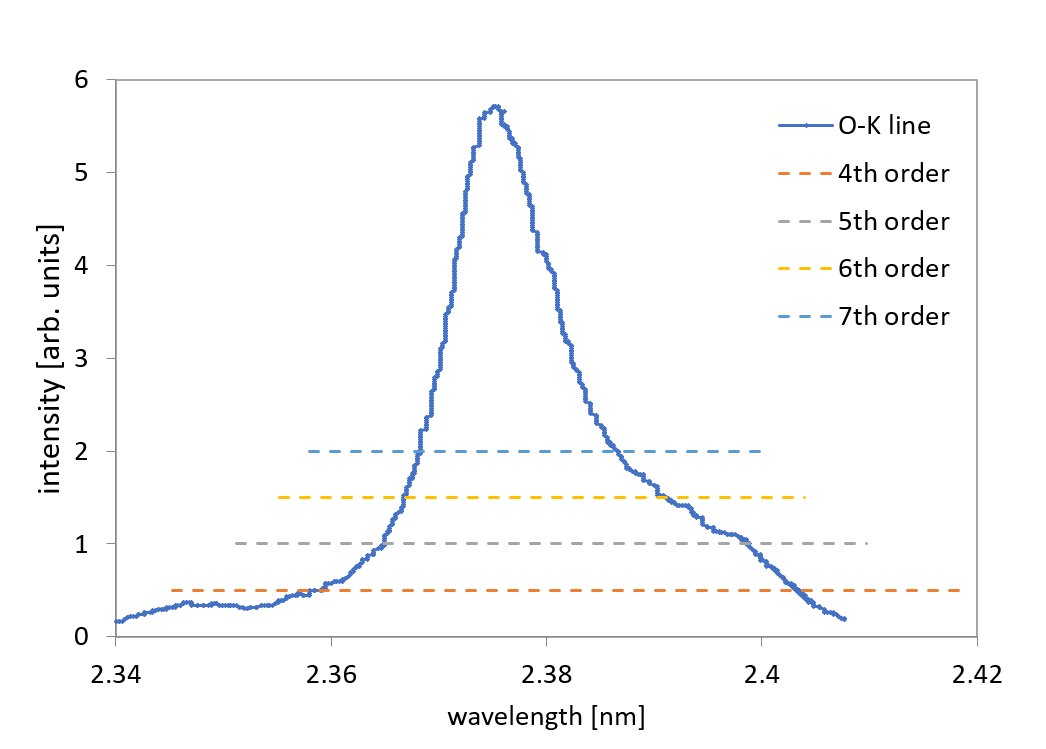}
   \includegraphics[height=7cm]{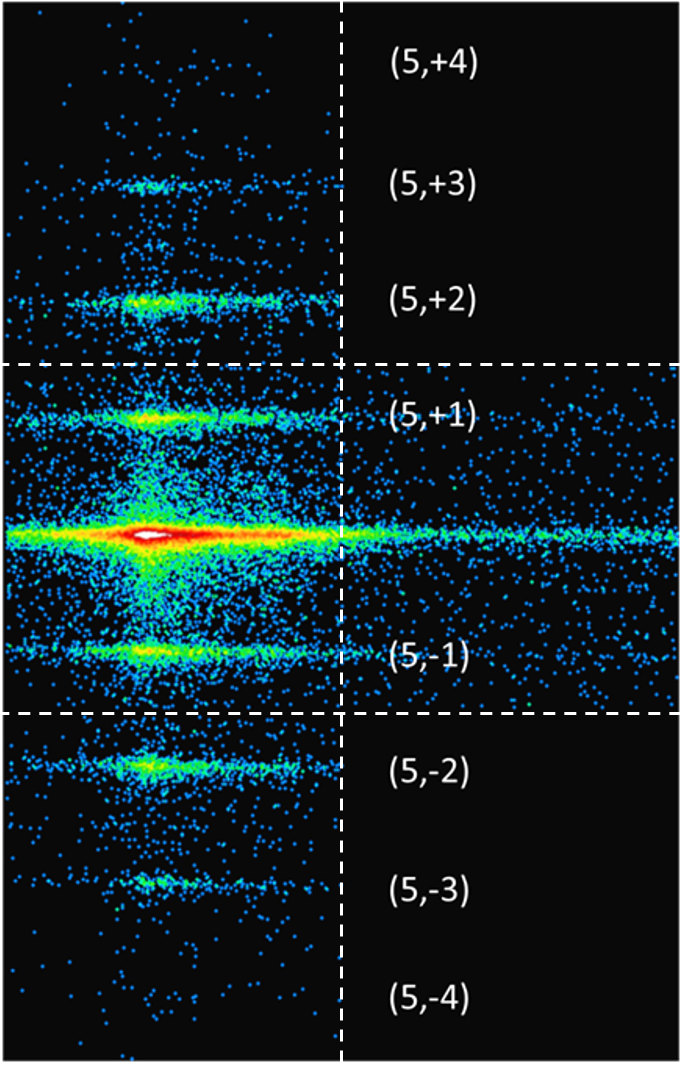}
   \end{tabular}
   \end{center}
\caption{Left: O-K$_{\alpha}$ spectrum from \citep{Remond}. The dashed lines show which part of the spectrum falls on a single CCD frame centered on orders 4 through 7.  Right: Mosaic of four CCD exposures for $5^{\mathrm {th}}$ order O-K$_{\alpha}$, showing the $(5,n)$ L1 diffraction orders with $n = 0, \pm 1,...,\pm 4$ (logarithmic intensity scale).  The central frame on the (horizontal) CAT grating dispersion axis only integrates over L1 orders $n = 0, \pm 1$. The frame on the right collects a small part of the O-K$_{\alpha}$ spectrum and clearly shows the underlying bremsstrahlung continuum.  For most gratings and diffraction orders we only collected the central image frame. \label{fig:O-Kspectrum}}
\end{figure}

The O-K$_{\alpha}$ transitions ($\lambda \sim 2.38$ nm, see Fig.~\ref{fig:O-Kspectrum}) lie near the middle of the Arcus bandpass and near hot plasma OVII and OVIII absorption lines expected in the continua from bright background AGN.  We used a silicon oxide target as our anode to create characteristic O-K x rays and to measure grating diffraction efficiency under Arcus-like conditions.  Here this means illuminating a grating at a graze angle of $\theta = 1.8$ degrees relative to the grating bars, centering the TRoPIC camera on a diffracted order (for orders 4-6), measuring the count rate over the full CCD image of $19.2 \times 19.2$ mm$^2$, and normalizing the count rate to the count rate of the beam incident upon the grating (``direct beam", measured with the gratings moved out of the way).  We call this ratio the diffraction efficiency of the grating, even though it does not just include the diffraction efficiency of the CAT grating bars, but also the effects of the presence of the opaque L1 and L2 support structures.

A number of aspects of this measurement require careful consideration:  

Avoiding pile-up in CCD pixels from the focused beam requires a large reduction in source flux and thus count rate.  We alleviated this problem slightly by moving the camera 200 mm out of focus, but we still had to reduce the count rate to 3-4 cts/sec over the selected energy range of the CCD (450-600 eV).  For reasonable counting statistics each exposure had to be at least 15 minutes long.

For the weakest orders of interest the count rates were up to 50 times lower than for the direct beam, which would lead to exceedingly long exposure times.  At the time of our experiments there was no beam monitor available, so we had to increase the source flux in steps when measuring the direct beam, the $0^{\mathrm {th}}$ order transmitted beam, and higher diffracted orders.  Orders 0 and 5 had to be measured at two different source settings each in order to be able to compare count rates between direct beam and $0^{\mathrm {th}}$ order, as well as $0^{\mathrm {th}}$ and $5^{\mathrm {th}}$ order.  Orders 4-7 were measured at the same source settings. ($7^{\mathrm {th}}$ order was included in our measurements since we could reach it with the CCD, even though the Arcus readouts will only collect orders 4-6 at O-K.)

Without a beam monitor we had to rely on the stability of the flux from the source. This introduces relative uncertainty on the order of $\sim 10$\%, but on isolated occasions we observed even greater variations in flux.

We operated the source at an acceleration voltage of 1.8 keV and without a Si filter.  Therefore the source spectrum consisted of the O-K spectrum on top of a broad bremsstrahlung continuum.
The CCD energy resolution is $\sim 70$-80 eV (FWHM).  The count rate for the direct beam and $0^{\mathrm {th}}$ order includes the sum of continuum and the O-K spectrum integrated over the selected energy range, and convolved with the CCD energy resolution.  For non-zero diffracted orders the continuum is dispersed in angle, and only a fraction of the continuum that is integrated over in the direct beam and the $0^{\mathrm {th}}$ order will land on the CCD.  We therefore cannot simply divide count rates for normalization as would be the case for monochromatic x rays (see ``count ratio" column in Table \ref{tab:Eff}), but have to model the source spectrum for proper normalization.

The L1 cross supports diffract weakly in the direction perpendicular to the CAT grating dispersion direction, i.e., along the CAT grating cross-dispersion (CD) direction.  When talking about L1 diffraction we label orders $(m,n)$, where $m$ is the order diffracted from the CAT grating bars, and $n$ is the order due to diffraction from the L1 mesh.  For O-K$_{\alpha}$ wavelengths only the $n = 0,\pm 1$ orders are caught by a single CCD frame (see Fig.~\ref{fig:O-Kspectrum}), and the count rate could be reduced by a few percent due to the missing counts from the $|n|>1$ orders.

Similarly, looking at Fig.~\ref{fig:O-Kspectrum} we see that the O-K spectrum is rather broad, and for $5^{\mathrm {th}}$ and higher orders exceeds the CCD width, which also leads to slightly reduced count rates when comparing to the direct beam.

\begin{figure}[ht!]
   \begin{center}
   \begin{tabular}{ c c } 
   \includegraphics[height=10cm]{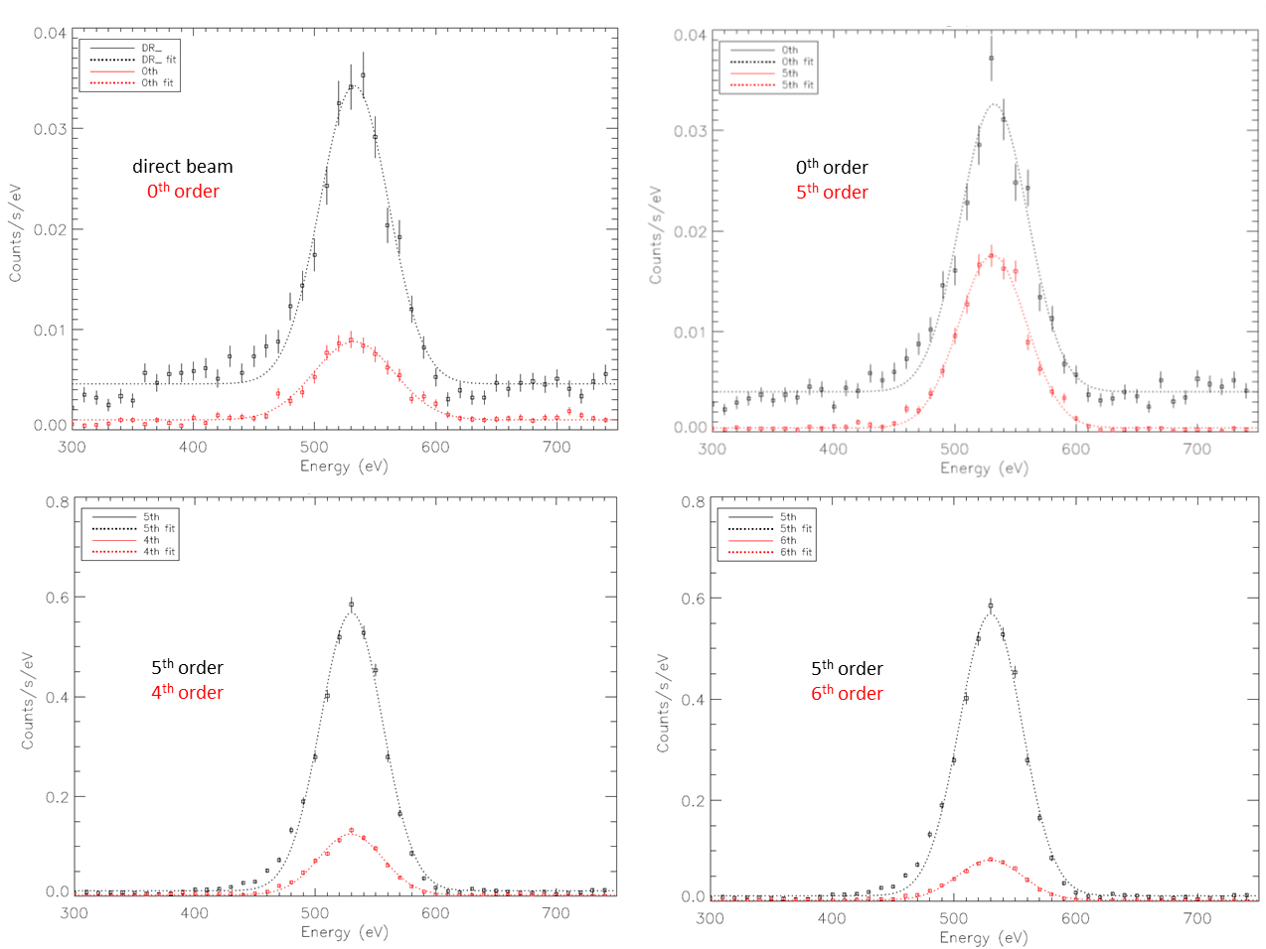}
   \end{tabular}
   \end{center}
\caption{Diffraction efficiency measurement: Counts/sec/eV of the O-K$_{\alpha}$ line as a function of energy measured by the TRoPIC camera for grating CNS1, using a SiO$_2$ anode.  Dotted lines are Gaussian fits to the data with a constant backgound.  The chosen source flux was lowest for the comparison of the direct beam (grating removed) and the $0^{\mathrm {th}}$ order (top left), higher for the comparison of $0^{\mathrm {th}}$ and $5^{\mathrm {th}}$ order (top right), and highest for comparison between the non-zero orders (only orders 4-6 are shown).  
\label{fig:norm}}
\end{figure}

In Fig.~\ref{fig:norm} we compare the count rates for grating CNS1 as a function of energy for the direct beam and orders 0 and 4-6.  Within each panel the source settings were the same.  We model the O-K$_{\alpha}$ line, which is experimentally convolved with the CCD energy resolution, as a Gaussian, and the continuum as a constant background.  We then calculate the fraction of the counts under the Gaussian relative to the total number of counts, and correct the total count rate for the exposure by this factor.  

The result is then used for relative scaling of count rates between exposures with different source settings.  The measurements for gratings SEG25 and SEG30 look qualitatively the same as Fig.~\ref{fig:norm}.  Results after correcting for the effects of the continuum and the width of the O-K spectrum are listed in bold in Table \ref{tab:Eff} (column ``continuum and frame size corrected").  Unfortunately we ran out of time to execute a complete set of measurements on grating CNS5.

\section{Comparison with synchrotron measurements}

Gratings CNS1, SEG25 and SEG30 have been previously measured at beamline 6.3.2 of the Advanced Light Source (ALS) at Lawrence Berkeley National Laboratory.  In order to compare these measurements with the above diffraction efficiency results from the PANTER setup a number of differences between the experiments need to be considered. They are mostly related to beam size, the range of structures and the areas illuminated by the beam, the spectral composition of the sources, and the angular range of the detectors.

\begin{figure}[ht!]
   \begin{center}
   \begin{tabular}{ c c } 
   \includegraphics[height=5cm]{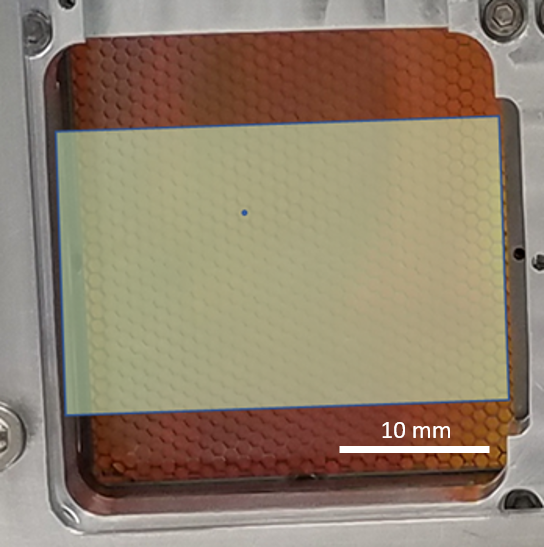}
   \end{tabular}
   \end{center}
\caption{Comparison of beam footprints overlayed on a photograph of the backside of grating SEG25. The large rectangle shows the approximate area ($\sim 625$ mm$^2$) illuminated by the SPO, sampling across hundreds of L2 hexagons. The small dot approximates the area of the synchrotron beam footprint, centered within a single hexagon. \label{fig:footprint}}
\end{figure}

\begin{figure*}
\gridline{
          \fig{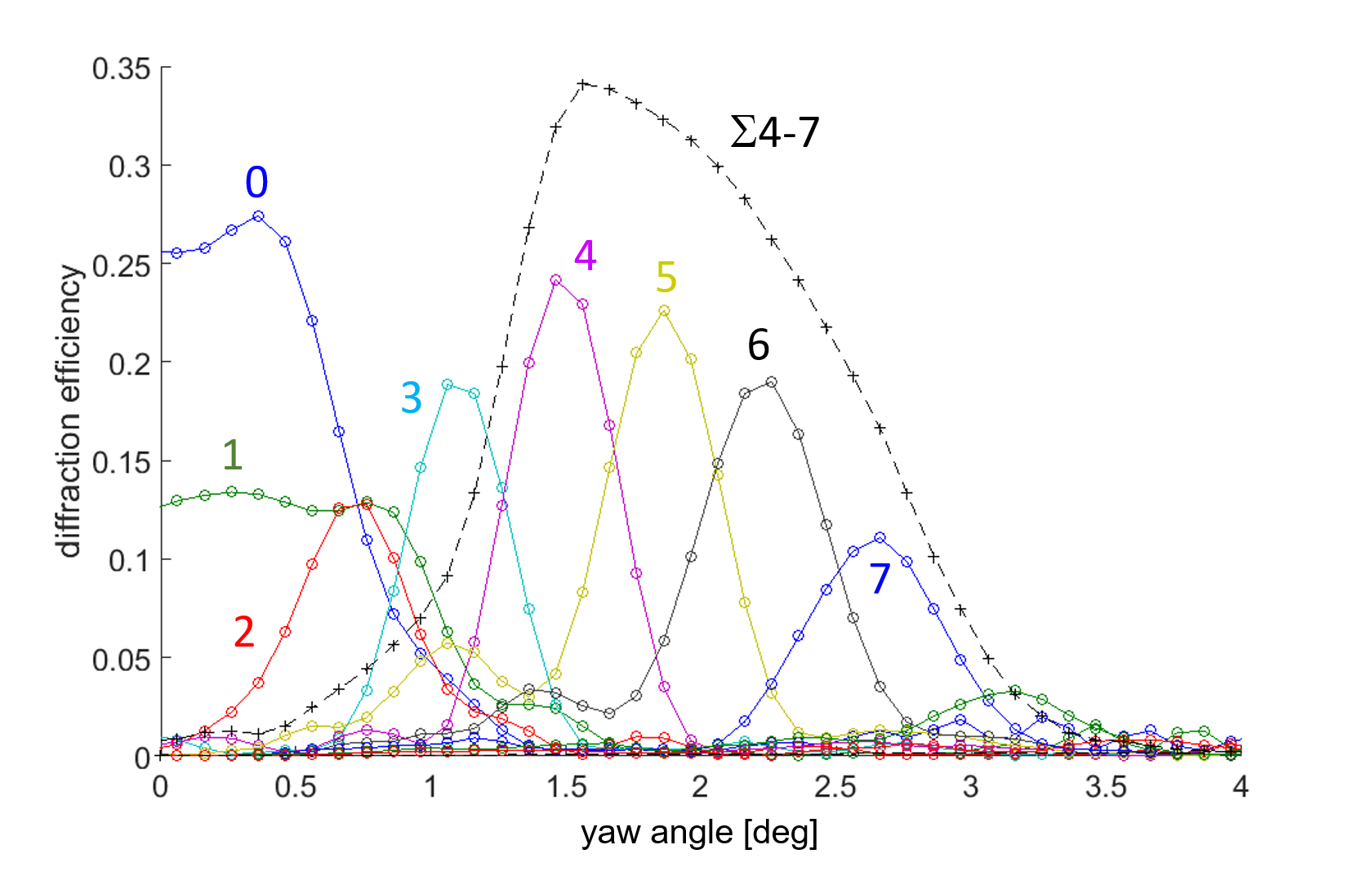}{0.55\textwidth}{(a)}
          \fig{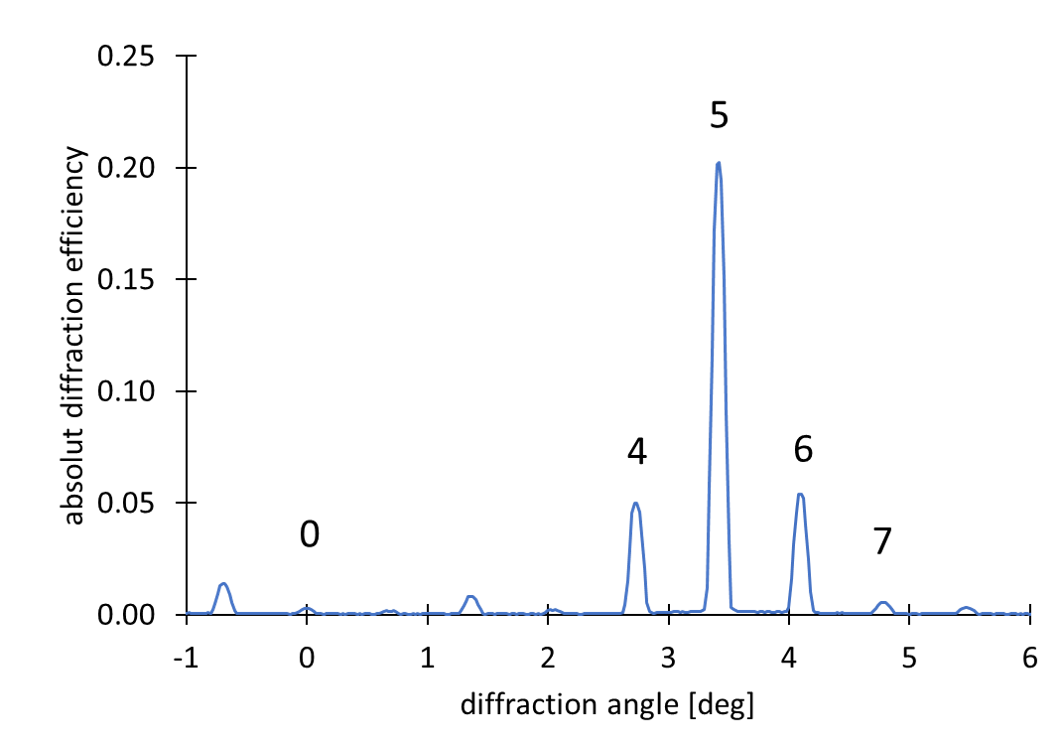}{0.45\textwidth}{(b)}
          }

\caption{Example synchrotron diffraction efficiency data from grating SEG25 at $\lambda = 2.38$ nm.  (a) Grating yaw scan for orders $m = 0$-7. The black dashed line is the sum of orders 4-7.  Since SEG25 is $\sim 5.5$ $\mu$m deep it is more efficient at slightly smaller angles than the Arcus design value of 1.8 degrees.  (b)  Detector scan at a fixed yaw angle of 1.8 degrees. Blazing is maximized for diffraction orders near twice the incidence angle onto the grating bar sidewalls (orders 4-6).  See also Fig.~\ref{fig:CAT_schematic}(a) and Eq.~\ref{ge}.  The synchrotron data includes absorption by the L1 mesh.
\label{fig:synch}}
\end{figure*}

At the ALS the collimated synchrotron beam was centered inside a single L2 hexagon, with the footprint of the beam being less than half of the area enclosed by the hexagon (see Fig.~\ref{fig:footprint}) and sampling over roughly 100 L1 mesh bars.  The wavelength of the x rays can easily be tuned with $\lambda/(\Delta \lambda) > 1000$.  As point detector a slit-covered photodiode (slit dimensions 0.5 mm in dispersion direction, 3 mm in CD direction) sits 230 mm from the grating and measures current which is proportional to the x-ray flux incident on the diode. For a typical scan we positioned the detector at the angle of the transmitted diffraction order of interest and rotated the grating in yaw.  Due to the relaxed alignment tolerances in the transmission geometry the angular movement of the order is negligible and the detector can remain at a fixed angle even when yawing by several degrees.  Since the beam is practically monochromatic the scan data is normalized simply through division by the diode current from the direct beam (measured at the same source settings, with the grating moved out of the way).  We tuned to a wavelength of 2.38 nm for comparison with the O-K$_{\alpha}$ spectrum.  Modeling shows that the variation of diffraction efficiency over the width of the O-K$_{\alpha}$ spectrum is negligible in this context.  See Fig.~\ref{fig:synch} for example data.

The diode integrates at least over L1 diffraction orders $n = 0,\pm 1,...,\pm 25$. Diffraction efficiency for L1 orders falls off very quickly with increasing $n$, i.e. only the first few L1 orders need to be considered, and the synchrotron data sums over all relevant L1 orders, while the PANTER CCD only collects orders $n = 0,\pm 1$ in a single frame.

We estimated the loss in count rate for a single PANTER CCD frame due to higher L1 diffraction orders falling outside the CCD by comparing the count rates in the individual L1 diffraction orders from mosaic images like Fig.~\ref{fig:O-Kspectrum}.  One can see that orders $n = \pm 4$ (and higher) can be neglected.  We found that orders $n = 0, \pm 1$ contained about 88-92\% of the total counts in orders $n = 0, \pm 1,...\pm 4$.  In the last column of Table \ref{tab:Eff} we apply the respective factor for each grating to the synchrotron data.

The O-K$_{\alpha}$ spectrum in Fig.~\ref{fig:O-Kspectrum} is broad, and part of it falls outside a single CCD frame for orders $m=4$ and higher.  The missed flux fraction increases with increasing order (less than 1\% for $4^{\mathrm {th}}$ order, about 5\% for $7^{\mathrm {th}}$ order).  We apply the appropriate frame size correction to each order in the second column of Table \ref{tab:Eff}.

At 2.38 nm wavelength both the L1 and the L2 support structures are opaque to x rays. Due to the different beam sizes, blockage from the L2 mesh only occurs for the PANTER data.  The dimensions of the L2 hexagons are on the scale of 0.1-1.0 mm and are lithographically defined with high precision.  We therefore simply calculate the L2 open area fraction from the L2 design parameters at 81\% and also consider shadowing at angled incidence (\citep{moritz_SPIE2019}) of 1.8 degrees to estimate total blockage from the L2 mesh at 20.7\%.  We apply this correction to the synchrotron data for comparison with PANTER data (column labeled ``after L2 correction" in Table \ref{tab:Eff}).  

Comparing the bold columns we see that the average diffraction efficiency performance over a large fraction of a grating in a slowly converging beam is similar to individual spot measurements under pencil beam illumination, and that we have not inadvertently selected peak performance spots for our synchrotron measurements.

\begin{table}
\begin{center}
\caption{Comparing diffraction efficiencies (sum over orders 4-7) from PANTER and synchrotron measurements.  
\label{tab:Eff} }
\begin{tabular}{l|cc|ccc}
Grating & $\Sigma_{4-7}$ & $\Sigma_{4-7}$ & $\Sigma_{4-7}$ & after L2 correction & after L1 diffraction \\
 & (count ratio)* &  (continuum and frame size corrected)* & synchrotron & (open area + shadowing) & correction \\
\hline
CNS1 & 0.204 & \textbf{0.283} & 0.312 & 0.247 & \textbf{0.219}  \\
SEG25 & 0.198 & \textbf{0.251} & 0.327 & 0.259 & \textbf{0.228}  \\
SEG30 & 0.203 & \textbf{0.237} & 0.350 & 0.278 & \textbf{0.247} \\
\hline
 \end{tabular}
\end{center}
{\small
   *The first two columns have an estimated relative uncertainty of at least 10\% due to variations in source flux over time.
   }
\vspace{-0.1in}
\end{table}

\section{comparison with model predictions}

\begin{figure}[ht!]
   \begin{center}
   \begin{tabular}{ c c } 
   \includegraphics[height=6cm]{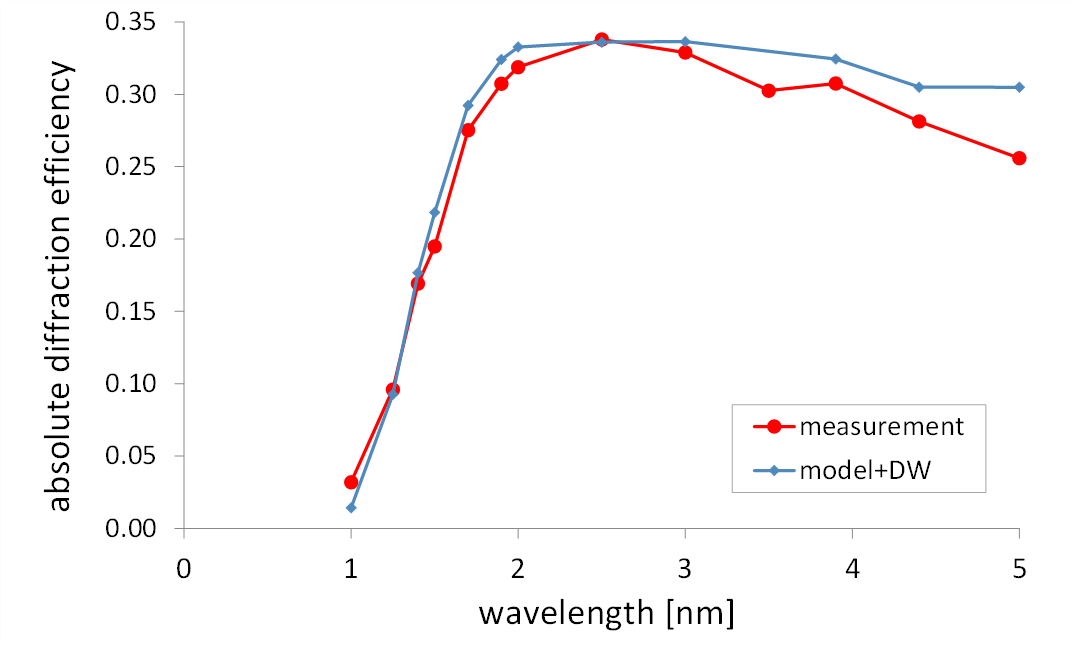}
   \end{tabular}
   \end{center}
\caption{Comparison of model efficiency (sum of blazed orders, multiplied by Debye-Waller-like roughness factor, for 82\% L1 open area fraction) with synchrotron spot measurements (sum of blazed orders) for a previous generation $32 \times 32$  mm$^2$ CAT grating (\citep{SPIE2017}).  The L2 mesh does not play a role in this plot. \label{fig:efficiency}}
\end{figure}

Previously, we have compared synchrotron measurements with RCWA model predictions for grating efficiency.  The RCWA model only considers the CAT grating bars, but since the synchrotron measurements integrate over many L1 diffraction orders we can take the impact of the L1 mesh into account simply by multiplying the model predictions with the open area fraction of the L1 mesh (\citep{AO2011}).  A heuristic Debye-Waller-(DW)-like roughness factor $\sigma_r$ is used to model grating imperfections (\citep{SPIE2020}). 
 The resulting model predictions provided a satisfactory match with synchrotron data taken at several wavelengths between 1 and 5 nm (see Fig.~\ref{fig:efficiency}).  We then calculated L2 blockage and shadowing and used the result as input for effective area predictions for Arcus.

The above PANTER measurements on almost fully SPO-illuminated Arcus-like gratings give more direct and realistic information about large-area grating performance in an Arcus-like configuration than synchrotron spot measurements.  Nevertheless, we need to close the loop between measurements and model-based effective area predictions.  The Arcus model assumes 60 nm wide and 4 $\mu$m deep CAT grating bars.  We sum the model diffraction efficiencies of orders 4-7 at an incidence angle of 1.8 degrees, and apply a roughness factor of $\sigma_r = 2$ nm.  

The design L1 open area fraction in the OPL photomask was 85\%.  However, the binary nature of the mask together with the applied lithography illumination resulted in optical proximity effects that made the L1 bars wider than desired and reduced the L1 open area fraction to 78\%.  In future masks we will either bias the mask toward thinner L1 bars or use phase shifting photomasks to obtain the design L1 bar width.  

After we apply the realized open area fractions for L1 and L2 and the L2 shadowing factor, the Arcus model predicts an effective diffraction efficiency of 22.9\%.  This is the equivalent number that underlies the effective area prediction for Arcus at $\lambda = 2.38$ nm.  As shown in column 2 of Table \ref{tab:Eff}, all three gratings performed in agreement with or in excess of this number.

\section{Comparison of relative roll angles from diffraction of visible light and x rays}

As mentioned in Section \ref{sec:exp}, we measured the initial roll angles of each grating using red light diffraction from the L1 mesh in air.  After the x-ray measurement campaign we calculated the relative roll angles between the four gratings from the centroids of diffracted x-ray orders ($7^{\mathrm {th}}$ order for O-K$_{\alpha}$ and $18^{\mathrm {th}}$ and $21^{\mathrm {st}}$ order for Al-K$_{\alpha}$) relative to $0^{\mathrm {th}}$ order, using camera pixel coordinates and camera stage positions during the respective measurements.

We found $\Delta U = -1.7\pm 0.4$ arcmin (SEG25), $28.4\pm 0.5$ arcmin (CNS1), and $-52.8\pm 0.6$ arcmin (CNS5) relative to SEG30.  These numbers agree with the laser-based measurements within uncertainties as expected due to the fixed angle between the CAT grating bars and the L1 cross-supports in the photomask used to pattern all the gratings.

\vspace{10mm}

\section{discussion}

We discuss the above results and their meaning for Arcus.

\subsection{Grating roll alignment}

Roll alignment using L1 diffraction was only done between grating membrane and facet frame.  It was not controlled when mounting the facets to the grating window.  In the future we plan to monitor facet-to-facet and facet-to-window roll alignment during mounting and to adjust it if necessary.  L1 red light diffraction roll measurements taken at PANTER agreed with roll results from CAT grating x-ray diffraction data, but were somewhat imprecise due to additional diffraction from the SPO.  Our lab-based L1 diffraction setup can be made more precise since there we have control over the optical design, can shape the laser beam, and do not need an SPO in the path.

\subsection{Resolving power}

As with a number of measurements of effective grating resolving power of previous CAT gratings, we found $R_G$ in the range of $\sim 7 \times 10^3$ to $1.3\times 10^4$. The differences compared to previous experiments are that here we illuminated up to 20 times the grating area, and that we showed for the first time that the combination of spectra from two separate gratings patterned with OPL and bonded to flight-like metal frames did not lead to a reduction in $R_G$.  This demonstrates that the fabricated grating facets have good uniformity and the same average grating period, at least to within $\Delta p = p/R_G = 17$ pm for SEG25 and SEG30.
The two deeper gratings seem to have slightly higher $R_G$.  We speculate that this might be due to the fact that the deeper grating bars suspended between deeper L1 cross supports are expected to be stiffer than the shallower versions, leading to smaller potential gap variations between grating bars.  However, within $3\sigma$ uncertainties we can not clearly distinguish between the four gratings.

\begin{figure}[ht!]
   \begin{center}
   \begin{tabular}{ c c } 
   \includegraphics[height=6cm]{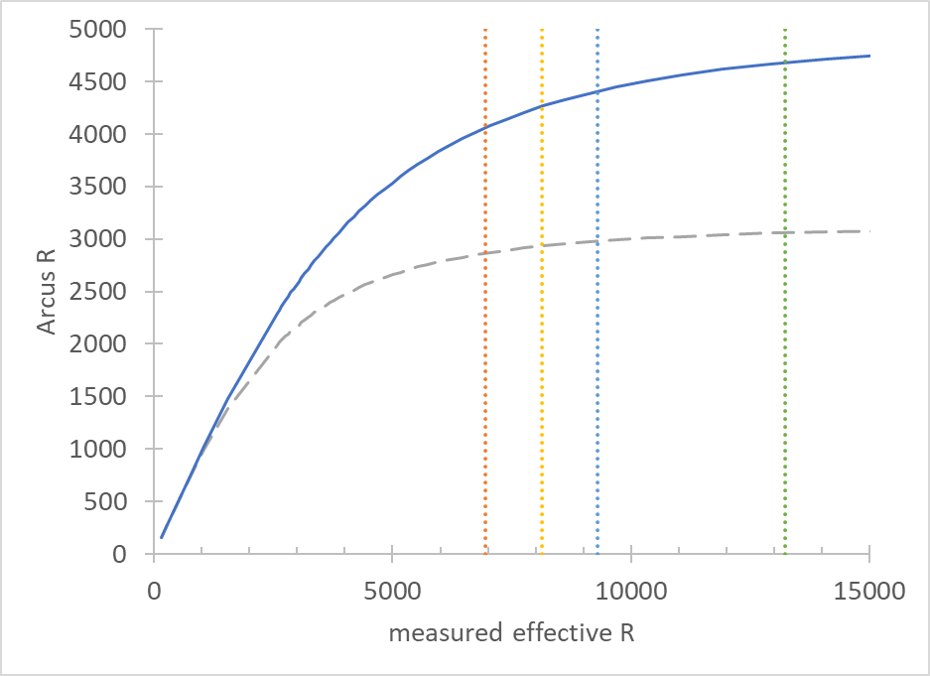}
   \end{tabular}
   \end{center}
\caption{Estimated Arcus resolving power $R$ as a function of measured effective grating resolving power $R_G$, assuming that all gratings have the same value of $R_G$.  The dashed curve assumes that all other PSF-contributing factors perform at the required level, while the solid curve assumes that all other factors perform at the level of current best estimates.  The dotted vertical lines show the $R_G$ best fit values for the gratings measured in $18^{\mathrm {th}}$ order.
\label{fig:errorbudget}}
\end{figure}

The broadening $\sigma_G$ conservatively assigned to each of the gratings in the estimation of $R_G$  depends, of course, on the width of the direct beam, $\sigma_{DB}$, and the validity of the assumption of a Gaussian direct beam profile.  However, as discussed in Sect.~\ref{meas} and shown in Fig.~\ref{fig:DB}, we need the sum of two Gaussians for a good fit.  We continued our conservative approach and used the smaller FWHM of 0.944 arcsec from the double Gaussian fit in Eq.~\ref{R}, which practically assigns any broadening from the direct beam tails to the gratings, leading to lower, more conservative $R_G$ values. 

The TRoPIC measurement of the direct beam gave a wider FWHM of $1.15\pm 0.06$ arcsec.  Since we did not perform sub-pixel raster scans as described elsewhere \citep{Dennerl_SPIE2012}, we do not obtain the maximum possible spatial resolution enhancement to a level comparable to PIXI.  Based on previous experience with event grade selections similar to the ones utilized in this work, we expect sub-pixel resolution of about 40 $\mu$m  at Al-K energies.  If we simply model detector spatial resolution as Gaussian broadening and extrapolate from 20 $\mu$m pixel PIXI data to 40 $\mu$m spatial resolution we would expect a FWHM of $\sim 1.09$ arcsec, which agrees with the measured result within uncertainties.

The FWHM of the PSF for a full Arcus OC has many contributors besides the gratings, such as the FWHM of the full SPO petal, various misalignments between all the different elements of an OC, and dynamic contributors, such as pointing reconstruction errors and jitter.  The Arcus team maintains error budgets for all relevant terms for two scenarios: required performance (RP) and current best performance estimates (CBE) for all terms.  The resulting PSF FWHM in the dispersion direction, together with the photon distribution into the collected diffraction orders, dictates the resolving power of an OC.  In Fig.~\ref{fig:errorbudget} we show the expected Arcus resolving power $R$ as a function of $R_G$, if all gratings in the array had the same value of $R_G$.  The dashed curve shows the case where all other elements perform at the required level, and the solid curve shows the case where all other elements perform at the level of CBEs.  Vertical lines show the best fit values for $R_G$ from the $18^{\mathrm {th}}$ order data of the four gratings.  As we can see, the requirement of $R \ge 2500$ can be met if all gratings deliver $R_G \ge 4100$ in the RP scenario, and $R_G \ge 2900$ in the CBE scenario.  Our measurements show that all four tested gratings safely exceed this performance by factors ranging from 1.7 to 4.6.
In the RP scenario the variation in measured $R_G$ values has very little impact on performance.  However, in the CBE case gains in $R_G$ can lead to more noticeable improvements in $R$.  In practice the roughly 500 gratings required for Arcus will have a distribution of $R_G$ values.  If the four gratings measured here were representative of such a distribution Arcus would easily meet its required (RP case) and goal (CBE case) performance.  Since we have never measured a CAT grating with $R_G < 6000$ we are confident that Arcus-like grating facets fabricated in the presented fashion will lead to a high-$R$ Arcus mission.

\subsection{Diffraction efficiency and effective area}

Over the last decade we have measured CAT grating diffraction efficiency at the ALS, and more recently also in our own x-ray facility (\citep{SPIE2018,Garner_SPIE2019}).  We have previously investigated grating uniformity by scanning gratings across a synchrotron beam (\citep{SPIE2015,SPIE2016,SPIE2017}), but this is a time-consuming process and does not fully mimic illumination in the Arcus-like converging beam of an SPO.  Here we have verified that our gratings perform in accordance with expectations under Arcus-like conditions and closed the loop between PANTER and synchrotron measurements and model expectations for effective area.

The only differences between this batch of gratings and the Arcus design are the wider L1 bars (1.1 instead of 0.75 $\mu$m) and the thicker device layer for gratings SEG25 and SEG30 ($\sim 5.5$ instead of 4.0 $\mu$m).  We previously fabricated $32\times 32$ mm$^2$ gratings with 0.5 $\mu$m wide L1 bars (\citep{SPIE2019}) and do not see any obstacles to reducing L1 bar width to the Arcus design value for future batches via OPL mask design adjustments.

In this work we only measured efficiency at a single characteristic line.  However, if we have reasonable knowledge of the realized geometric grating bar parameters ($d,b$), for example from SEM imaging,  we can extrapolate performance to other wavelengths using RCWA.  While extrapolation can provide useful estimates we still need to measure performance at other wavelengths to obtain precise results.  In addition to the ALS beamline, we can now also do this at the MIT x-ray polarimetry beamline (\citep{Garner_SPIE2020}), using a range of available anodes.

Diffraction efficiency, and thus effective area, can be increased by making deeper gratings and using them at slightly smaller blaze angles (\citep{SPIE2021} - see also Fig.~\ref{fig:synch}).  Since this would also shift the blaze peak toward lower orders it would potentially reduce $R_G$.  How this might affect Arcus performance is best investigated via ray tracing.  Deeper gratings are more difficult to fabricate, so there are multiple trade-offs to be considered.

Other obvious means of increasing effective area are reductions in the widths of grating bars and L1 and L2 structures.  The trade-offs here are manufacturability and mechanical strength.  Reductions in all three structures are under active investigation.

CAT grating facets with 0.75 and 0.90 $\mu$m wide L1 bars previously have undergone thermal cycling in vacuum and vibration testing without negative impact on their x-ray performance (\citep{SPIE2017}).  These tests still have to be repeated for the current generation of Arcus CAT grating prototypes and for populated grating windows to confirm the expected ruggedness.

\section{summary and conclusions}

We have fabricated four Arcus-sized membranes from three different 200 mm SOI wafers using patterning, deposition, and etch tools that are compatible with volume production.  The silicon membranes were epoxied to metal facet frames with the average CAT grating bar angle nominally adjusted parallel to the frame normal.  The grating facets were mounted to a grating window without metrology feedback.  The grating window was placed in the converging beam of an SPO at the PANTER x-ray facility.  The average grating bar angles for each facet were measured to be parallel to each other with a standard deviation of 1.8 arcmin, meeting the error budget, and demonstrating that we can adjust for bar tilt resulting from the DRIE step.  

We measured the Al-K$_{\alpha}$ doublet in $18^{\mathrm {th}}$ and $21^{\mathrm {st}}$ orders and derived effective grating resolving powers in the range of $R_G \sim 6.9 \times 10^3$ to $1.3 \times 10^4$, exceeding the Arcus requirement by factors of $\sim$ 2-4. For the two gratings that were aligned well to each other in roll we demonstrated $R_G = 1.3^{+\infty}_{-0.5} \times 10^4$ from the combined spectrum of the simultaneously illuminated gratings, showing that with proper alignment between two different gratings the spectral resolving power does not degrade.

Diffraction efficiency was measured for three gratings at O-K$_{\alpha}$ wavelengths, the center of the Arcus bandpass, with near 80\% illumination of the area of a grating and at the Arcus design yaw angle of 1.8 deg. Comparison with previous synchrotron diffraction efficiency spot measurements at 2.38 nm wavelength showed agreement within measurement uncertainties.  The performance of all three measured gratings met or exceeded the value used for the Arcus effective area model at this wavelength.

We have shown that we can manufacture multiple flight-like Arcus gratings with volume production compatible techniques and tools, and that they meet or exceed the required performance specifications for a successful Arcus mission.

\begin{acknowledgments}

This work was supported by NASA grants 80NSSC19K0335 and 80NSSC250K0780 and the MIT Kavli Institute for Astrophysics and Space Research. It made use of the Shared Experimental
Facilities at MIT supported in part by the MRSEC Program of the National Science Foundation under award
number DMR 1419807.  We appreciate facility support from the MIT Nanostructures Lab and MIT.nano.  This research also used resources of the Advanced Light Source (beamline 6.3.2), a U.S. DOE Office of Science User Facility under contract no.~DE-AC02-05CH11231.
Part of this work has been supported by the European Union’s Horizon 2020 Programme under the AHEAD2020 project (grant agreement n.~871158).
We thank Mallory Whalen for machining and transportation help.

\end{acknowledgments}

%

\vspace{5mm}






\bibliography{CAT_references}{}
\bibliographystyle{aasjournal}



\end{document}